\documentclass[pra,aps,twocolumn,groupedaddress]{revtex4-1}
\usepackage{color}
\usepackage{graphicx}
\usepackage{hyperref}
\usepackage{amsmath}
\usepackage{latexsym}
\usepackage{amssymb}
\usepackage{mathrsfs}
\usepackage{mathtools}
\usepackage{layout}
\usepackage{verbatim}
\usepackage{bm}
\usepackage{amsfonts,epsfig}

\begin{document}
\title{Rydberg Quantum Gates Free from Blockade Error}

%\date{\today}
\author{Xiao-Feng Shi}
\affiliation{School of Physics and Optoelectronic Engineering, Xidian University, Xi'an 710071, China}

\begin{abstract}
  Accurate quantum gates are basic elements for building quantum computers. There has been great interest in designing quantum logic gates by using blockade effect of Rydberg atoms recently.  The fidelity and operation speed of these gates, however, are fundamentally limited by an intrinsic blockade error. Here we propose a type of quantum gates, which are based on Rydberg blockade effect, yet free from any blockade error. In contrast to the ``blocking'' method in previous schemes, we use Rydberg energy shift to realize a rational generalized Rabi frequency so that a $\pi$ phase for one input state of the gate emerges. This leads to an accurate Rydberg quantum logic gate that can operate on a 0.1-$\mu s$ timescale or faster because it works by a Rabi frequency which is comparable to the blockade shift. 

\end{abstract}

\maketitle
\section{introduction}
Quantum computation holds fascinating features that are not shared by a classical computer. This originates from the nature of quantum bits~(qubits) exhibited via exotic phenomena such as quantum superposition and interference. Even so, a quantum computer also needs the very basic logic gates to operate.  To design a quantum logic gate, a rich variety of systems, such as semiconductor quantum dots~\cite{Awschalom2013}, superconducting circuits~\cite{Chuang2013}, photonic qubits~\cite{Pan2012}, and atomic ions~\cite{Ballance2016,Gaebler2016} have been investigated extensively. Recently, neutral atoms excited to high-lying states, usually called Rydberg states~\cite{Gallagh2005}, have also inspired intensive laboratory interest~\cite{Isenhower2010,Zhang2010,Maller2015,Wilk2010,Jau2015} for the demonstration of potential quantum computation based on neutral atoms~\cite{PhysRevLett.85.2208,Lukin2001,Muller2009,Saffman2010,Saffman2016}.

Most quantum logic gate experiments~\cite{Isenhower2010,Zhang2010,Maller2015} with Rydberg atoms worked in the blockade regime~\cite{Urban2009,Gaetan2009,Dudin2012,Peyronel2012,PhysRevLett.115.013001,Beguin2013} where a blockade shift $V$ approximately forbids a Rabi cycle for a certain input state, upon which a $\pi$ phase shift is subsequently induced~\cite{PhysRevLett.85.2208}. This blockade regime, however, is intrinsically accompanied with a blockade error. Such an error is proportional to the square of a factor $\mathscr{M}$, where $\mathscr{M}\equiv\hbar\Omega/V$ and $\Omega$ is the (effective) Rabi frequency for exciting the relevant Rydberg state~\cite{Saffman2005}~[$h~(\hbar)$ denotes the (reduced) Planck's constant]. Decreasing $\Omega$ and hence $\mathscr{M}$ may reduce the blockade error, but inevitably increases the gate time and hence the error due to decay of Rydberg states, which is another important error. By using an experimentally accessible $V$, a gate with fidelity error of about $10^{-3}$ was theoretically predicted~\cite{Zhang2012}. Such a fundamental limit originates from the necessary condition of the gate: $\mathscr{M}$ should be much smaller than 1 so that the gate can work properly; unfortunately, this imposes that the gate time should be much larger than $h/V$, accompanied by a  significant probability of Rydberg state decay. This seems to suggest that only exceedingly large $V/h$~(perhaps on the GHz scale, see Ref.~\onlinecite{Theis2016}) may help to reduce the operation time and hence the fidelity error of a conventional Rydberg gate toward the goal of scalable quantum computation~\cite{Preskill1998}.

Here we propose a type of Rydberg-interaction-based two-qubit quantum gate protocols free from any blockade error, leaving the decay of the Rydberg states and population leakage out of the computational basis as the only source of intrinsic error. Specifically, our gate protocol requires $\mathscr{M}\sim1$, in sharp contrast to the condition $\mathscr{M}\ll1$ of previous schemes. Consequently, it becomes possible to apply much larger $\Omega$ compared with those in Refs.~\cite{Isenhower2010,Wilk2010,Zhang2010,Maller2015,Jau2015}. Because a larger $\Omega$ renders shorter gate operation time and reduced error due to Rydberg state decay, our protocols provide a route to build a high-fidelity two-qubit quantum gate with currently achievable MHz-scale Rydberg blockade shift and laser Rabi frequencies, while shortening the gate operation times significantly. As shown later on, a $C_Z$ gate according to our protocol can be accomplished with an error smaller than $10^{-4}$ with both $V/h$ and $\Omega/2\pi$ in the order of $10$~MHz.

\begin{figure}
\includegraphics[width=3.4in]
{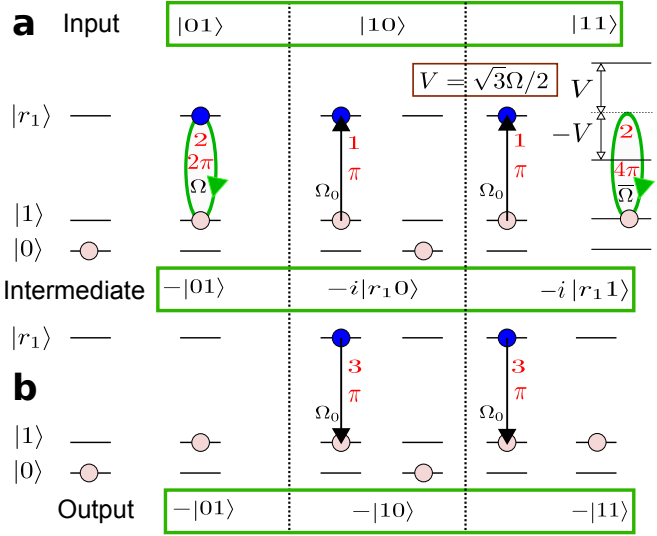}
 \caption{A 3-pulse sequence for a $C_{\text{Z}}$ gate protocol with dipole-dipole interaction. Subsequent application of the pulses in (a) and (b) will perform the $C_{\text{Z}}$ gate. The number accompanying each arrowed line/curve denotes the $k$th pulse, while $A\pi$, where $A=1,2$ or $4$, specifies the pulse area. Optical pumping occurs only for the logic state $|1\rangle$. Here the blockade shift $V$ of the state $|r_1r_1\rangle$ satisfies $V=\sqrt3\hbar\Omega/2$, so that the generalized Rabi frequency satisfies $\overline{\Omega}=2\Omega$. Such a condition guarantees that Pulse-2 simultaneously maps the states $|01\rangle$ and $|r_11\rangle$ back to themselves, respectively.   \label{fig001} }
\end{figure}

\begin{figure}
\includegraphics[width=3.4in]
{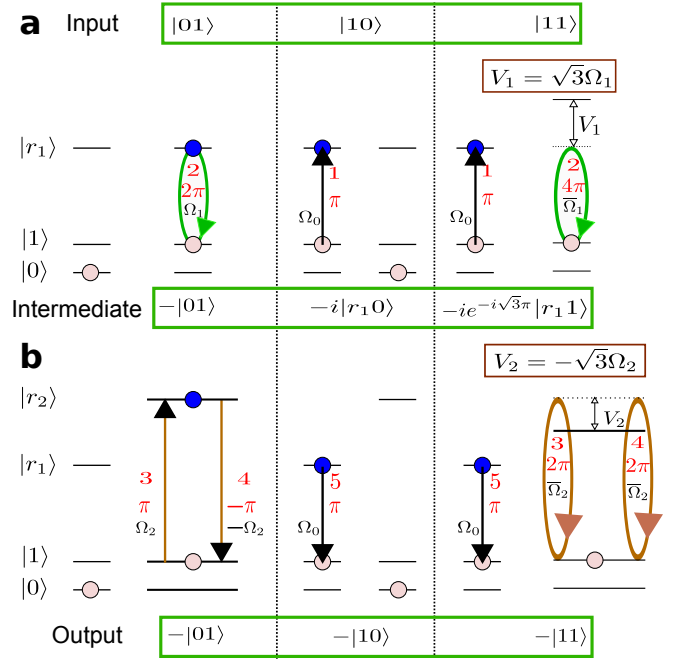}
 \caption{A 5-pulse sequence for a $C_{\text{Z}}$ gate protocol by using van der Waals interaction. Here the blockade shift $V_1~(V_2)$ of the state $|r_1r_1\rangle~(|r_1r_2\rangle)$ satisfies $V_1=\sqrt3\hbar\Omega_1$~($V_2=-\sqrt3\hbar\Omega_2$), so that the generalized Rabi frequency satisfies $\overline{\Omega}_k=2\Omega_k$, where $k=1$ or $2$.  \label{fig002} }
\end{figure}

Such $C_{\text{Z}}$ gate protocols are schematically shown in Figs.~\ref{fig001} and~\ref{fig002}, with two typical interaction types, namely, first-order dipolar interaction and second-order van der Waals interaction, respectively. The $C_{\text{Z}}$ gate here performs the state transformation $\{|00\rangle,|01\rangle,|10\rangle,|11\rangle\}\mapsto \{|00\rangle,-|01\rangle,-|10\rangle,-|11\rangle\}$, where $|\alpha\beta\rangle\equiv |\alpha\rangle\otimes|\beta\rangle$ is a two-qubit product state, and $|\alpha(\beta)\rangle=|0(1)\rangle$ is an s-orbital hyperfine ground state of an alkali-metal atom of two different hyperfine levels $F=(I\pm 1/2)\hbar$, where $I$, the nuclear spin quantum number, is equal to $3/2(7/2)$ for $^{87}$Rb~($^{133}$Cs), a frequently employed isotope in experiments~\cite{Isenhower2010,Wilk2010,Zhang2010,Maller2015,Jau2015}. Here the left~(right) digit inside the ket denotes the state of the control~(target) qubit. The basic idea of our protocols is that the Rydberg blockade $V$ is used to create a rational generalized Rabi frequency
\begin{equation}
  \overline{\Omega}\equiv \sqrt{\Omega^2+\eta (V/\hbar)^2};~\overline{\Omega}=2\Omega, \label{equation1}
  \end{equation}
which is in sharp contrast to traditional methods where $V$ is used to block a Rabi oscillation. 
$\overline{\Omega}$ determines the time evolution of the lower state involved in the (partial) Rabi cycle between two states,
where $\Omega$ is the single-atom Rabi oscillation frequency in the absence of $V$, and $\eta=1(4)$ if we use van der Waals~(direct dipolar) interaction between two Rydberg atoms, as detailed later on. Our protocol makes advantage of the fact that when $\overline{\Omega}=2\Omega$, a $\pi$ pulse for $\Omega$ is equivalent to a $2\pi$ pulse for $\overline{\Omega}$, where $\Omega~(\overline{\Omega})$ determines the state evolution of a certain intermediate state $|01\rangle(|r_11\rangle)$ that is transformed from the input state $|01\rangle~(|11\rangle)$. When $\overline{\Omega}=2\Omega$, applying a $2\pi$ pulse of $\Omega$ upon the target qubit will simultaneously map the two different intermediate states, $|01\rangle$ and $|r_11\rangle$, back to themselves, with a phase accumulation of $\pi$ and $0~[$or $- \pi V/(\hbar\overline{\Omega})]$, respectively. Such a peculiar phase can then be used for building a two-qubit quantum gate. So, the blockade shift in our protocol is not necessarily large compared with $\Omega$, as a requirement in the conventional quantum entanglement experiments with Rydberg atoms~\cite{Isenhower2010,Wilk2010,Zhang2010,Maller2015,Jau2015}. Most importantly, we do not use $V$ to block any Rabi oscillation, hence there is no blockade error. Since the blockade error is a major factor limiting the gate performance ~\cite{Saffman2016}, our protocol may provide a versatile platform for generating high-efficiency quantum logic gates with neutral atoms.

\section{A 3-pulse protocol of a $C_{\text{Z}}$ gate by using first-order dipolar interaction}\label{sec02}
%{\it A 3-pulse protocol of a $C_{\text{Z}}$ gate by using first-order dipolar interaction.--}
In detail, our gate protocol by using first-order dipolar interaction, shown in Fig.~\ref{fig001}, is designed with 3 pulses. In Fig.~\ref{fig001}, each laser excitation of the atoms is resonant with the transition $|1\rangle\leftrightarrow|r_1\rangle$, where $|r_1\rangle$ refers to a Rydberg state of the control or target qubit. When Rabi frequencies up to several $100$~MHz are used, the coupling between the laser and the atom in the state $|0\rangle$ is largely detuned because of the large hyperfine splitting $\omega_{g}$ between the two logic states $|0\rangle$ and $|1\rangle$, which is about $h\times6.8~(9.2)$~GHz for $^{87}$Rb~($^{133}$Cs). For a first approximation, the input state $|00\rangle$ does not change during the 3-pulse sequence, and it will be sufficient to study the other three input states as in Fig.~\ref{fig001}. The rotation error ignored in this approximation will be analyzed later on. 

Before the explanation of the protocol in Fig.~\ref{fig001}, it will be useful to first briefly review the dipolar interaction of three degenerate states $\{|pf\rangle,|r_1r_1\rangle,|fp\rangle\}$~\cite{Ravets2014}. Since D-orbital states allow relatively easier~(compared with S-orbital states) optical access by two-photon transitions, we choose an example of $|r_1\rangle=|nD_{5/2},m_J=5/2\rangle$, so that $|p\rangle=|(n+2)P_{3/2},m_J=3/2\rangle$ and $|f\rangle=|(n-2)F_{7/2},m_J=7/2\rangle$ for $^{87}$Rb. Notice that for $|r_1\rangle=|nD_{3/2},m_J=3/2\rangle$, $|p\rangle=|(n+2)P_{1/2},m_J=1/2\rangle$ and $|f\rangle=|(n-2)F_{5/2},m_J=5/2\rangle$ of $^{87}$Rb, the examples of $n=58$ and $59$ have been experimentally studied~\cite{Gaetan2009,Ravets2014}. Here we choose $J=5/2$, instead of $J=3/2$, so that the residual excitation of the other fine level can be avoided when we only employ right-hand polarized lasers throughout the gate sequence. Dipolar interaction induces resonant transitions $|r_1r_1\rangle\leftrightarrow\{|pf\rangle,|fp\rangle\}$ with the following diagonalized Hamiltonian~\cite{Gaetan2009,Ravets2014}
\begin{eqnarray}
 \mathcal{H}_{\text{dd}}&=& V(|S_+\rangle\langle S_+|-|S_-\rangle\langle S_-|),\nonumber
\end{eqnarray}
where $V$ represents a coupling strength from dipole-dipole interaction, $|S_\pm\rangle = (|pf\rangle \pm\sqrt{2}|r_1r_1\rangle+|fp\rangle )/2$, while the dark eigenstate is $|S_0\rangle = (|pf\rangle -|fp\rangle )/\sqrt2$. More details could be found in Appendix~\ref{appendixA}. With the dipolar interaction at hand, below we study the state evolution for the input states, $\{|01\rangle,|10\rangle,|11\rangle\}$, of the $C_Z$ gate. For the sake of convenience, the $k$th pulse will be termed as Pulse-$k$ below, where $k=1,2,\cdots$.

We first look at Fig.~\ref{fig001}(a) about the first two pulses. Pulse-1 upon the control qubit has an area of $\pi$ with a Rabi frequency $\Omega_0$, exciting the control qubit through the transition $|1\rangle\leftrightarrow|r_1\rangle$. Pulse-2 upon the target qubit has a Rabi frequency $\Omega$ for the transition $|1\rangle\leftrightarrow|r_1\rangle$, and lasts for a time of $2\pi/\Omega$. It is easy to show that a $\pi$ phase will be added to the input state $|01\rangle$ upon the completion of Pulse-2, but not so obvious about what will happen for $|r_11\rangle$. To study this, the system Hamiltonian during Pulse-2
\begin{eqnarray}
\mathcal{ H}^{(\text{d})}&=& \mathcal{H}_{\text{dd}}+\hbar\Omega(|r_11\rangle\langle r_1r_1|+\text{h.c.})/2\nonumber
\end{eqnarray}
 can be diagonalized as $ \mathcal{H}^{(\text{d})}=\sum_{\alpha=\pm} \alpha \hbar\overline{\Omega} |v_\alpha\rangle\langle v_\alpha|/2$, where $\overline{\Omega} = \sqrt{\Omega^2+4(V/\hbar)^2}$ and $|v_\pm\rangle$ are eigenvectors~[see Eq.~(\ref{v0pm})]. Starting from the initial state at the beginning of Pulse-2, i.e., $|\psi(0)\rangle=-i|r_11\rangle = -i\sum_{k=0,\pm}a_k|v_k\rangle$, the wavefunction evolves as~[see Appendix~\ref{appendixA} for details]
\begin{eqnarray}
 |\psi(t)\rangle &\propto& V |v_0 \rangle +\Omega(e^{-it\overline{\Omega}/2} |v_+ \rangle +e^{it\overline{\Omega}/2}|v_- \rangle )/2^{3/2} .\nonumber
\end{eqnarray}%$|\psi(t)\rangle=-ie^{-iV_1t/2}\sum_{k=1}^2a_ke^{\mp i\overline{\Omega}_1 /2}|v_k\rangle $.
In order to map the state $|r_11\rangle$ back to it again at the end of Pulse-2, we choose, as an example, $V=\sqrt3\hbar \Omega/2$, so that $\overline{\Omega}=2\Omega$. This results of $|\psi(2\pi/\Omega)\rangle=-i|r_11\rangle $ at the end of Pulse-2, and the state evolution induced by the first two pulses can be summarized as,
\begin{eqnarray}
 \{|01\rangle,|10\rangle,|11\rangle\}\mapsto-\{|01\rangle,i|r_10\rangle,i|r_11\rangle\}.\label{eq03}
\end{eqnarray}
Pulse-3 in Fig.~\ref{fig001}(b) has a laser excitation scheme that is identical to Pulse-1, so as to perform a mapping which is inverse to that realized by Pulse-1,
\begin{eqnarray}
 -\{|01\rangle,i|r_10\rangle,i|r_11\rangle\}&\mapsto& - \{|01\rangle,|10\rangle,|11\rangle\},\nonumber
\end{eqnarray}
completing the $C_{\text{Z}}$ gate with the understanding that the input state $|00\rangle$ is intact during the pulse sequence. The whole sequence of laser excitation is summarized in Fig.~\ref{fig001}.

\section{A 5-pulse protocol of a $C_{\text{Z}}$ gate by using second-order van der Waals interaction}\label{sec03}
%    {\it A 5-pulse protocol of a $C_{\text{Z}}$ gate by using second-order van der Waals interaction.--}
    Below we discuss how to implement a $C_{\text{Z}}$ gate protocol free from blockade error by the commonly used van der Waals interaction, as shown in Fig.~\ref{fig002} with 5 optical pulses. The reason that we use different excitation schemes is because the first-order dipole-dipole and second-order van der Waals interactions induce different effects, as seen from Fig.~\ref{fig001}(a) and Fig.~\ref{fig002}(a). For this reason, two types of Rydberg states $|r_1\rangle$ and $|r_2\rangle$ appear in the protocol of Fig.~\ref{fig002}, in order to have two interactions $V_1$ and $V_2$ of different signs. Here $V_1$ and $V_2$ are the energy shifts of $|r_1r_1\rangle$ and $|r_1r_2\rangle$, respectively. Details for the calculation below could be found in Appendix~\ref{phasevdWI}.

In Fig.~\ref{fig002}(a), Pulse-1 upon the control qubit has an area of $\pi$ with Rabi frequency $\Omega_0$, exciting the control qubit through a transition $|1\rangle\leftrightarrow|r_1\rangle$. Pulse-2 upon the target qubit has a Rabi frequency $\Omega_1$ for the transition $|1\rangle\leftrightarrow|r_1\rangle$, and lasts for a duration of $2\pi/\Omega_1$. For the input state $|11\rangle$ during Pulse-2, the Hamiltonian is
\begin{eqnarray}
  \mathcal{H}^{(\text{v})}&=&V_1|r_1r_1\rangle\langle r_1r_1| +\hbar  \Omega_1(|r_11\rangle\langle r_1r_1|+\text{h.c.})/2 .\nonumber%\label{Hamil2}
\end{eqnarray}%written in the ordered basis $\{|r_1r_1\rangle, |r_11\rangle\}$.
$ \mathcal{H}^{(\text{v})}$ can be diagonalized as $\sum_{\alpha=\pm} \epsilon_\alpha |v_\alpha\rangle\langle v_\alpha|$, where $\epsilon_\pm = (V_1 \pm \hbar \overline{\Omega}_1)/2$ is the eigenvalue of the eigenvector $|v_\pm\rangle$, and $\overline{\Omega}_1\equiv\sqrt{\Omega_1^2+(V_1/\hbar)^2}$~[see Eq.~(\ref{v0pmvdwI})]. With the two eigenstates $|v_\pm\rangle$, we recast the initial state $|\psi(0)\rangle=-i|r_11\rangle$ at the beginning of Pulse-2 into $|\psi(0)\rangle= -i\sum_{k=1}^2a_k|v_k\rangle$, so that its subsequent time evolution can be derived as
\begin{eqnarray} |\psi(t)\rangle=-ie^{-iV_1t/2\hbar}\sum_{k=1}^2a_ke^{\mp i\overline{\Omega}_1 t/2}|v_k\rangle.\nonumber
\end{eqnarray}%$|\psi(t)\rangle=-ie^{-iV_1t/2}\sum_{k=1}^2a_ke^{\mp i\overline{\Omega}_1 /2}|v_k\rangle $.
In order to map the state $|r_11\rangle$ back to it again at the end of Pulse-2, we choose $V_1=\sqrt3 \hbar\Omega_1,$
%\begin{eqnarray} V_1=\sqrt3 \Omega_1,\end{eqnarray}
so that $\overline{\Omega}_1=2\Omega_1$. This results of $|\psi(2\pi/\Omega_1)\rangle=-ie^{-i\sqrt3\pi}|r_11\rangle $ at the end of Pulse-2, thus the state evolution induced by the first two pulses is,
\begin{eqnarray}
 \{|01\rangle,|10\rangle,|11\rangle\}\mapsto-\{|01\rangle,i|r_10\rangle,ie^{-i\sqrt3\pi}|r_11\rangle\}.\label{eq06}
\end{eqnarray}
Compared with Eq.~(\ref{eq03}), the state $|r_11\rangle$ in Eq.~(\ref{eq06}) has an extra phase term $e^{-i\sqrt3\pi}$ due to the difference of the interaction mechanisms for the state $|r_1r_1\rangle$~[see Figs.~\ref{fig001}(a) and~\ref{fig002}(a)]. 

The next three pulses are schematically shown in Fig.~\ref{fig002}(b). Pulse-3 upon the target qubit resonantly couples $|1\rangle$ and a Rydberg state $|r_2\rangle$, which is different from $|r_1\rangle$. We choose $|r_2\rangle\neq |r_1\rangle$ so as to have a negative blockade shift $V_2$~(when $V_1>0$), which is possible when the principal quantum numbers of the states $|r_1\rangle$ and $|r_2\rangle$ are different if we use s or p-orbital states~(see Appendix~\ref{S-vdWI} or Ref.~\cite{Walker2008}). By applying Pulse-3 with a Rabi frequency $\Omega_2=|V_2|/(\sqrt3\hbar)$ and duration of $\pi/\Omega_2$, a similar calculation that leads to Eq.~(\ref{eq06}) gives the following state transformation during Pulse-3
\begin{eqnarray}
 - \{|01\rangle,i|r_10\rangle,ie^{-i\sqrt3\pi}|r_11\rangle\}&\mapsto&\{i|0r_2\rangle,-i|r_10\rangle, \nonumber\\
  &&ie^{-i\sqrt3\pi/2}|r_11\rangle\}.\nonumber%\label{eq07}
\end{eqnarray}
Similar to Pulse-2, here the peculiar feature that the same Pulse-3 drives the two states, $|01\rangle$ and $|r_11\rangle$, completely to the two respective states $|0r_2\rangle$ and $|r_11\rangle$, is because of the relation $\overline{\Omega}_2=2\Omega_2$: a $\pi$ pulse for the transition $|01\rangle\leftrightarrow|0r_2\rangle$ is equal to a $2\pi$ pulse for the transition $|r_11\rangle\leftrightarrow|r_1r_2\rangle$, although the latter one is an incomplete Rabi process in the sense that the state $|r_1r_2\rangle$ is never fully populated.

The laser in Pulse-4 is designed to have the same central wavelength and intensity with those of Pulse-3, except of a $\pi$ phase difference. As a result, Pulse-4 drives the transition $|1\rangle\leftrightarrow|r_2\rangle$ with a Rabi frequency $-\Omega_2$. Similar calculation as used in Eq.~(\ref{eq06}) gives the following state transformation for Pulse-4,
\begin{eqnarray}
i\{|0r_2\rangle,-|r_10\rangle,e^{-i\sqrt3\pi/2}|r_11\rangle\}&\mapsto& -\{|01\rangle,i|r_10\rangle,i|r_11\rangle\}.\nonumber\label{keystep}
\end{eqnarray}
From the last two equations above, one finds that the change of the Rabi frequency from $\Omega_2$ to $-\Omega_2$ preserves the generalized Rabi frequency $\overline\Omega_2$. This is also an important feature for the $C_{\text{Z}}$ gate protocol here. 

Finally, we apply Pulse-5 which has the same physical property of Pulse-1, to complete the transformation of the $C_{\text{Z}}$ gate,
\begin{eqnarray}
 -\{|01\rangle,i|r_10\rangle,i|r_11\rangle\}&\mapsto& - \{|01\rangle,|10\rangle,|11\rangle\}.\nonumber\label{eq09}
\end{eqnarray}
Comparing the two protocols above, one finds that the latter one has two pulses, i.e., Pulse-3 and Pulse-4, which are absent in the first protocol. This is because of an extra phase term in Eq.~(\ref{eq06}) compared with Eq.~(\ref{eq03}), and the two extra pulses in Fig.~\ref{fig002} are designed to eliminate that phase term. In other words, we use three pulses, i.e., Pulse-2, Pulse-3 and Pulse-4 to realize Eq.~(\ref{eq03}) with van der Waals interaction, while only one pulse is enough for dipolar interaction.

\begin{figure}
\includegraphics[width=3in]
{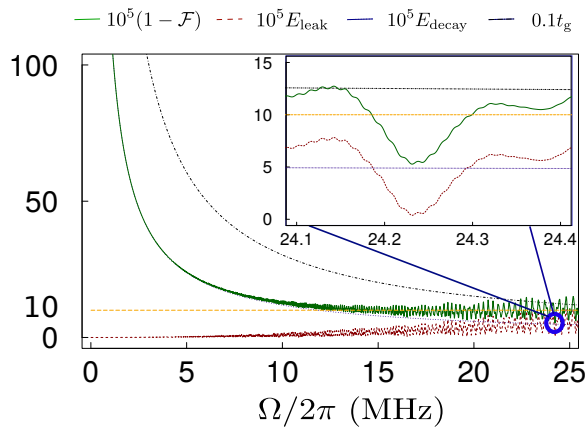}
 \caption{ Performance quality of the gate by using van der Waals interaction. Plotted are decay error $E_{\text{decay}}$, leakage error $E_{\text{leak}}$ and the total fidelity error $1-\mathcal{F}$, scaled by $10^5$. The lifetimes of the Rydberg states are estimated with $T=4$~K, the $C_6$ coefficients for the states $|r_1r_1\rangle$ and $|r_1r_2\rangle$ are $289$ and $-281$~(unit: $h\times$GHz$~\mu m^6$), respectively. Here $\Omega_j=\Omega$, $j=0,1$, and $\Omega_2=\Omega_1|V_2/V_1|$. The dash-dotted curve denotes the scaled gate time $t_{\text{g}}/10$. The circle at the right bottom denotes a case with $\Omega/2\pi=24.23$~MHz, $t_{\text{g}}=125$~ns, and $1-\mathcal{F}=5.28\times10^{-5}$. The enlargement of this circle is shown in the inset.     \label{fig03} }
\end{figure}

\section{Gate performance}
    %    {\it Gate performance.--}
        Below we study the performance of our gate protocol characterized by its fidelity error $1-\mathcal{F}$ and operation time $t_{\text{g}}$. For our protocols, there are two sources of intrinsic error for the gate fidelity, i.e., the Rydberg state decay and the population leakage to nearby unwanted transitions, while errors caused by atomic motion due to Rydberg state interaction can be neglected~[see the analysis below Eq.~(\ref{deltaV}) of Appendix~\ref{App0D}].  As can be numerically studied by assuming cooling atoms to their motional ground state in optical traps~\cite{Kaufman2012}, the position fluctuation of the qubits will also increase our gate fidelity error by one order of magnitude~(see Appendix~\ref{fluctuationError}), but it is not fundamental and in principle can be gradually removed with improved technology~\cite{Reiserer2013,Thompson2013,Kaufman2014,Lester2015}. By assuming ground-state cooling and an optical trap with a depth of $20$~mK, we may have a gate error of about $6\times 10^{-4}$~[see Fig.~\ref{p2} in Appendix~\ref{fluctuationError}].

The Rydberg state decay induces an error, $E_{\text{decay}}$, that is proportional to the time for the state to be in the Rydberg state~\cite{Saffman2005, Zhang2012}~(see Appendix~\ref{error2edorder01}), while the leakage error, $E_{\text{leak}}$, can be calculated numerically by taking all the important leaking channels into account~(see Appendix~\ref{error2edorder02}). As an example, we show in Fig.~\ref{fig03} the gate performance with our protocol using van der Waals interaction~(similar results can be found by using dipolar interaction, see Appendix~\ref{error1order}), where $|r_k\rangle=|n_kp_{3/2},m_J=3/2,m_I=3/2\rangle$, $k=1,2$, and $(n_1,n_2)=(80,90)$. The inset of Fig.~\ref{fig03} shows that it is possible to have a gate with $(t_{\text{g}},1-\mathcal{F})=(125$~ns, $5.3\times10^{-5})$ when $\Omega/2\pi=24.23$~MHz, denoted by the blue circle. The corresponding van der Waals interaction there is about $h\times40$~MHz, which is within current technical availability since a Rydberg-Rydberg interaction in the range of $h\times(5,50)$~MHz has been demonstrated~\cite{Isenhower2010,Wilk2010,Zhang2010,Maller2015,Jau2015,Gaetan2009}. Also, the Rydberg interaction by a direct excitation to a $np_{3/2}$ state, where $n>80$, has been realized~\cite{Hankin2014}. This means that with conventional square pulses, and Rydberg interactions~($/\hbar$) and Rabi frequencies  in the order of $10\times2\pi$~MHz, it is possible to build a two-qubit Rydberg quantum gate with fidelity larger than $0.9999$. Finally, our gate operation time in the order of $0.1\mu s$ compares favorably to that of a conventional Rydberg gate.

\section{conclusions}
To conclude, we have shown that based on the blockade shift of two Rydberg atoms, it is possible to realize a two-qubit controlled gate that is free from any blockade error. A $\pi$ phase for one of the four input states, as necessary for the gate, arises from a generalized Rabi cycle between a single-Rydberg state and a two-Rydberg state.  We have analyzed realization of this gate by using two $^{87}$Rb atoms for qubits, and found that our gate fidelity can be larger than $0.9999$  with both laser Rabi frequencies and blockade shift in the order of $10\times2\pi$~MHz.

 \section*{ACKNOWLEDGMENTS}
The author acknowledges support from the Fundamental Research Funds for the Central Universities and the 111 Project (B17035), and thanks T. A. B. Kennedy and Yan Lu for useful discussions.

\appendix{}
\section{Phase accumulation in a detuned Rabi cycle I: first-order dipole interaction}\label{appendixA}
Here we provide additional information on the theory of designing $C_Z$ gates free from any blockade error, especially about the emergence of the $\pi$ phase which is essential for adequate modeling of the system.  We would like to choose $^{87}$Rb as an example. The qubit states $|0\rangle$ and $|1\rangle$ are $|5S_{1/2},F=1,m_f=1\rangle$ and $|5S_{1/2},F=2,m_f=2\rangle$, respectively. Identities like $\hbar$ will not be written out explicitly. 

In this appendix we derive how a phase accumulates in a detuned Rabi cycle between a lower state and other three dipole-coupled upper states, which can be identified with the gate protocol using resonant dipole interaction in Sec.~\ref{sec02}. Using external electric field one can tune the energy levels, so that three states $|p\rangle=|(n+2)P_{1/2},m_J=1/2\rangle$, $|r_1\rangle\equiv|d\rangle=|nD_{3/2},m_J=3/2\rangle$, and $|f\rangle=|(n-2)F_{5/2},m_J=5/2\rangle$ of a $^{87}$Rb atom become exact degenerate with each other~\cite{Ravets2014}. The dipole interaction between two atoms, each of whom is excited to $|d\rangle$, was experimentally studied in Refs.~\cite{Gaetan2009,Ravets2014} with $n=58$ and $59$, respectively. As shown in Ref.~\cite{Gaetan2009}, the dipole interaction couples the three two-atom states $\{|pf\rangle,|dd\rangle,|fp\rangle\}$ through the following Hamiltonian,
\begin{eqnarray}
 H_{\text{dd}}&=& \left( \begin{array}{ccc}
0     &V_{\text{dd}} &0 \\
V_{\text{dd}} &0  &V_{\text{dd}}\\
0     &V_{\text{dd}} &0
    \end{array} \right), \label{Hdddefine}
\end{eqnarray}
which can be diagonalized with its three eigenstates
\begin{eqnarray}
|S_+\rangle &=& (|pf\rangle +\sqrt{2}|dd\rangle+|fp\rangle )/2,\nonumber\\
|S_0\rangle &=& (|pf\rangle -|fp\rangle )/\sqrt2,\nonumber\\
|S_-\rangle &=& (|pf\rangle -\sqrt{2}|dd\rangle+|fp\rangle )/2,\nonumber
\end{eqnarray}
and their respective eigenenergies $\sqrt2V_{\text{dd}},0$ and $-\sqrt2V_{\text{dd}}$. Below we set $\sqrt2V_{\text{dd}}\equiv \Delta$ for convenience.

In the gate protocol of Sec.~\ref{sec02} by using dipole interaction, we consider the following Hamiltonian written in the ordered basis $\{ |d1\rangle, |S_+\rangle,|S_-\rangle \}$ and define $\overline\Omega\equiv \Omega/(2\sqrt2)$, where $\Omega$ is the Rabi frequency between $|1\rangle$ and $|d\rangle$,
\begin{eqnarray}
 H_1&=& \left( \begin{array}{ccc}
     0&\overline\Omega&-\overline\Omega \\
     \overline\Omega&\Delta& 0\\
     -\overline\Omega&0& -\Delta
    \end{array} \right).\nonumber
\end{eqnarray}
Here $|S_0\rangle$ does not enter into $H_1$ because it is not coupled by the optical lasers. The signs of the two Rabi frequencies above differ because the coefficients of the component $|dd\rangle$ in $|S_\pm\rangle$ have different signs.

\begin{figure}
\includegraphics[width=3.4in]
{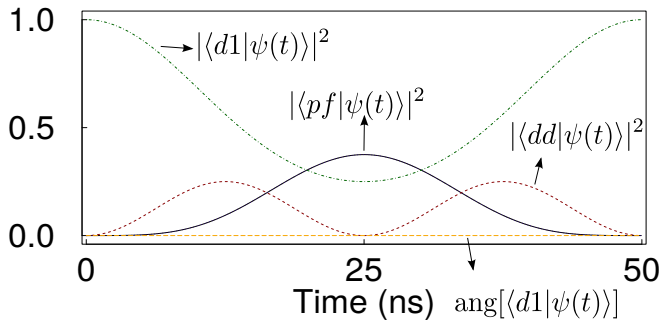}
 \caption{State evolution during Pulse-2 for the input state $|11\rangle$ of the $C_{\text{Z}}$ gate protocol using dipolar interaction. The initial state is $|d1\rangle$. Here $|\langle fp|\psi(t)\rangle|^2 =|\langle pf|\psi(t)\rangle|^2 $ and ang$[\langle d1|\psi(t)\rangle]$ is the argument of the component $|d1\rangle$ in the wavefunction. We use $\Omega/2\pi=20$~MHz here, while the strength of the dipolar interaction is determined by Eq.~(\ref{eqOmeV}). \label{S01} }
\end{figure}

The Hamiltonian $H_1$ can be diagonalized as
\begin{eqnarray}
  H_1&=& \sum_{\alpha=0,\pm} \epsilon_\alpha |v_\alpha\rangle\langle v_\alpha|.\nonumber
\end{eqnarray}
Here
\begin{eqnarray}
  \epsilon_\pm &=&\pm \overline{\epsilon},~\epsilon_0=0,\nonumber\\
  |v_- \rangle &=& \left[ 2\overline\Omega |d1\rangle + (\Delta-\overline{\epsilon}) |S_+\rangle+ (\Delta+\overline{\epsilon}) |S_-\rangle\right]/(2\overline{\epsilon}),\nonumber\\
  |v_+ \rangle &=& \left[ 2\overline\Omega |d1\rangle + (\Delta+\overline{\epsilon}) |S_+\rangle+ (\Delta-\overline{\epsilon}) |S_-\rangle\right]/(2\overline{\epsilon}),\nonumber\\
  |v_0 \rangle &=& \left[ \Delta |d1\rangle -\bar \Omega (|S_+\rangle+ |S_-\rangle)\right]/\overline{\epsilon},\label{v0pm}
\end{eqnarray}
where
\begin{eqnarray}
 \overline{\epsilon}&=&\sqrt{2\overline\Omega^2+\Delta^2} = \sqrt{\Omega^2/4+\Delta^2}.\nonumber
\end{eqnarray}
The inverse transformations give
\begin{eqnarray}
  |d1\rangle &=&\frac{\Delta |v_0 \rangle +\overline\Omega( |v_+ \rangle +|v_- \rangle )  }{\overline{\epsilon} } ,\nonumber
\end{eqnarray}
which means that for an initial state of $|\psi(0)\rangle=|d1\rangle$, the state evolves as
\begin{eqnarray}
  |\psi(t)\rangle &=&\frac{\Delta |v_0 \rangle +\overline\Omega(e^{-it\epsilon_+} |v_+ \rangle +e^{-it\epsilon_-}|v_- \rangle )  }{\overline{\epsilon} }.\nonumber
\end{eqnarray}
When the condition 
\begin{eqnarray}
\Omega=2\Delta/\sqrt{3} ,\label{eqOmeV}
\end{eqnarray}
is satisfied, we have
\begin{eqnarray}
 \overline{\epsilon}&=&\Omega.\nonumber
\end{eqnarray}
 Starting from $|01\rangle$, one Rabi cycle with a time
\begin{eqnarray}
t_{2\pi}=2\pi/ \Omega,\nonumber
\end{eqnarray}
for the transition $|01\rangle\leftrightarrow|0d\rangle$ is equivalent to two Rabi cycles for the transition $|d1\rangle\leftrightarrow|dd\rangle$, so that
\begin{eqnarray}
  |\psi(t_{2\pi})\rangle &= &\frac{\Delta |v_0 \rangle +\overline\Omega(e^{-2i\pi} |v_+ \rangle +e^{2i\pi}|v_- \rangle )  }{\overline{\epsilon} }=|d1\rangle.\nonumber
\end{eqnarray}
This means that by application of a $2\pi$ pulse of $\Omega$, $\{|01\rangle,|d1\rangle\}\mapsto \{-|01\rangle,|d1\rangle\}$, which means that the phase accumulations for the two input states $|01\rangle$ and $|11\rangle$ during Pulse-2 differ. This is the key step for realizing a $C_{\text{Z}}$ gate with dipole interaction. As a numerical test, the populations on each component and the argument of the component $|d1\rangle$ are shown in Fig.~\ref{S01} for the input state $|11\rangle$, which agrees with the analysis above.

At the beginning of this appendix, we discussed of realizing the dipolar interaction by the Rydberg states $|p\rangle=|(n+2)P_{1/2},m_J=1/2\rangle$, $|d\rangle=|nD_{3/2},m_J=3/2\rangle$, and $|f\rangle=|(n-2)F_{5/2},m_J=5/2\rangle$ for the discussion above. However, the fine structure splittings of the $p, d$ and $f$-orbital Rydberg states are quite small. Among them, the fine splitting is biggest for $p$-orbital states, and even this $p$-orbital splitting is smaller than $200$~MHz when $n>80$. For the experiment in Ref.~\cite{Ravets2014}, the level $|d\rangle=|59D_{3/2},m_J=3/2\rangle$ is lower than $|59D_{5/2},m_J=3/2\rangle$ by only about $20$~MHz. This means that excitation of both fine levels are possible with a MHz scale Rabi frequency. While it is not an issue for the phenomenon demonstration in~\cite{Gaetan2009,Ravets2014}, it is useful for us to choose another setting that would avoid this leakage: If we instead use all right-hand polarized laser beams, then starting from the ground state $|1\rangle \equiv |5S_{1/2}, F=2, m_f=2\rangle$ and via the intermediate state $|5P_{3/2}, F=3, m_f=3\rangle$, only the state $|nD_{5/2},m_J=5/2, m_I=3/2\rangle$ can be coupled. This can avoid any population leakage to the $D_{3/2}$ manifold. As for the question of whether we can have F\"orster resonance for such a choice, it was shown in Ref.~\cite{Walker2008} that the process $nd_{5/2}+nd_{5/2}\mapsto (n+2)p_{3/2}+(n-2)f$ can be near resonance even without externally applied field in certain cases of $n$. So, it is possible to realize the dipolar-interaction protocol analyzed above with the $D_{5/2}$ state.

\begin{figure}
\includegraphics[width=3.4in]
{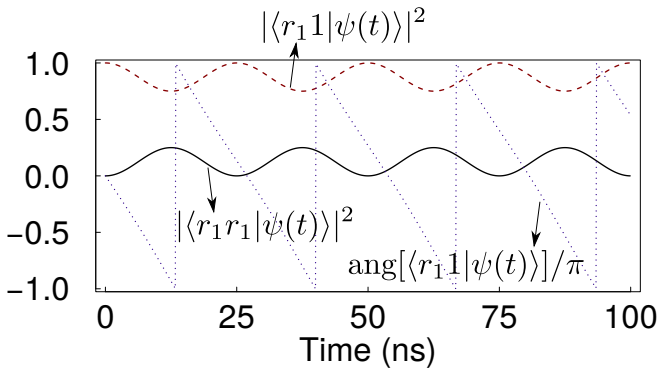}
 \caption{State evolution during Pulse-2 for the input state $|11\rangle$ of the $C_{\text{Z}}$ gate protocol using van der Waals interaction. The initial state is $|r_11\rangle$. Here ang$[\langle r_11|\psi(t)\rangle]$ is the argument of the component $|r_11\rangle$ in the wavefunction. We use $\Omega/2\pi=20$~MHz here, while the strength of the interaction is determined by Eq.~(\ref{relationOm1}). \label{S02} }
\end{figure}

%========================================================
%case 2
%=======================================================
\section{Phase accumulation in a detuned Rabi cycle II: second-order van der Waals interaction}\label{phasevdWI}
Below we study the phase accumulation in a detuned Rabi cycle for a two-level system, which can be identified with the gate protocol using van der Waals interaction of Sec.~\ref{sec03}. In this appendix, $V_1$ is the van der Waals interaction of the state $ |r_1r_1\rangle$. Consider the following Hamiltonian written in the ordered basis $\{ |r_1r_1\rangle, |r_11\rangle\}$,
\begin{eqnarray}
 H_{2}&=& \left( \begin{array}{cc}
     V_1&\Omega_1/2\\
    \Omega_1/2&0
    \end{array} \right).\nonumber%\label{H2defined}
\end{eqnarray}
The following calculation is valid for any system as long as the Hamiltonian takes the form above.

The Hamiltonian above can be diagonalized as
\begin{eqnarray}
  H_{2}&=& \sum_{\alpha=\pm} \epsilon_\alpha |v_\alpha\rangle\langle v_\alpha|,\nonumber
\end{eqnarray}
where
\begin{eqnarray}
  \epsilon_\pm &=& (V_1 \pm \sqrt{\Omega_1^2+V_1^2})/2,\nonumber\\
  |v_\pm \rangle &=& \left(\frac{\Omega_1}{2} |r_11\rangle + \epsilon_\pm |r_1r_1\rangle\right)/N_\pm,\nonumber\\
  N_\pm  &=& \sqrt{\Omega_1^2/4 + \epsilon_\pm^2}.\label{v0pmvdwI}
\end{eqnarray}
The inverse transformation gives
\begin{eqnarray}
  |r_11\rangle &=&\frac{2}{\Omega_1}\frac{\epsilon_-N_+ |v_+ \rangle - \epsilon_+N_- |v_- \rangle  }{ \epsilon_-  - \epsilon_+ } ,\nonumber\\
  |r_1r_1\rangle &=&\frac{N_+ |v_+ \rangle - N_- |v_- \rangle  }{ \epsilon_+  - \epsilon_- } ,\nonumber
\end{eqnarray}
which means that for an initial state of $|\psi(0)\rangle=|r_11\rangle$, the state evolves as
\begin{eqnarray}
  |\psi(t)\rangle &=&\frac{2}{\Omega_1}\frac{\epsilon_-N_+e^{-it\epsilon_+} |v_+ \rangle - \epsilon_+N_-e^{-it\epsilon_-} |v_- \rangle  }{ \epsilon_-  - \epsilon_+ } .\nonumber
\end{eqnarray}
Notice that a new characteristic frequency arises
\begin{eqnarray}
 \overline{\Omega}_1 \equiv \sqrt{\Omega_1^2+V_1^2},\nonumber
\end{eqnarray}
which determines the ground-state (partial) Rabi oscillation. Starting from $|r_11\rangle$, one Rabi cycle with a time
\begin{eqnarray}
t_{2\pi}=2\pi/ \overline{\Omega}_1,\nonumber
\end{eqnarray}
will simply induce the following phase change,
\begin{eqnarray}
  |\psi(t_{2\pi})\rangle &= &e^{-i\pi(1+V_1/\overline{\Omega}_1)}|r_11\rangle. \nonumber
\end{eqnarray}
For the gate protocol described in Sec.~\ref{sec02}, it will be useful if we impose a condition
\begin{eqnarray}
\overline{\Omega}_1 =2\Omega_1, \label{relationOm1}
\end{eqnarray}
so that by application of Pulse-2 in Fig.~\ref{fig002}, the states $|01\rangle$ and $|r_11\rangle$ will return to themselves, respectively. For Pulse-2 of the gate protocol in Sec.~\ref{sec03}, the populations on each component and the argument of the component $|r_11\rangle$ are shown in Fig.~\ref{S02} for the input state $|11\rangle$, which agrees with the analysis above.

From the result above, one finds that replacement of $\Omega_1$ by $-\Omega_1$ will not change the result, but change of $V_1$ to $-V_1$ will give nontrivial result. This peculiar feature will help one to understand how our gate by van der Waals interaction works.

\section{van der Waals interaction of two atoms in the state $|n_1p_{3/2};n_2p_{3/2}\rangle$}\label{S-vdWI}
Two-atom Rydberg blockade by a direct laser excitation from ground state to a p-orbital state with principal quantum number $n=84$ was ever demonstrated in Ref.~\cite{Hankin2014}. Thus it is practical to use p-orbital states for our gate protocol through van der Waals interaction. In this appendix we consider the interaction of two atoms, one of whom is in the state $| r_{\text{A}}\rangle = |n_{\text{A}}~^2P_{\frac{3}{2}}, m_J =  3/2\rangle $, and the other in $| r_{\text{B}}\rangle = |n_{\text{B}}~^2P_{\frac{3}{2}}, m_J = 3/2 \rangle $ and angle $\theta=0$, i.e., the two-atom separation axis coincides with the quantization axis. We consider the following nine channels for the dipole-dipole interaction, each characterized by its energy defect $\delta_k$, where $k=1,2,\cdots,9$,
\begin{eqnarray}
\delta_1(n_a,n_b) & = &  E(n_ad_{\frac{5}{2}}) + E(n_bd_{\frac{5}{2}}) -  E(n_{\text{A}} p_{\frac{3}{2}})-  E(n_{\text{B}} p_{\frac{3}{2}}),\nonumber\\
\delta_2(n_a,n_b) & = &  E(n_ad_{\frac{5}{2}}) + E(n_bd_{\frac{3}{2}}) -  E(n_{\text{A}} p_{\frac{3}{2}})-  E(n_{\text{B}} p_{\frac{3}{2}}),\nonumber\\
\delta_3(n_a,n_b) & = &  E(n_ad_{\frac{3}{2}}) + E(n_bd_{\frac{5}{2}}) -  E(n_{\text{A}} p_{\frac{3}{2}})-  E(n_{\text{B}} p_{\frac{3}{2}}),\nonumber\\
\delta_4(n_a,n_b) & = &  E(n_ad_{\frac{5}{2}}) + E(n_bs_{\frac{1}{2}}) -  E(n_{\text{A}} p_{\frac{3}{2}})-  E(n_{\text{B}} p_{\frac{3}{2}}),\nonumber\\
\delta_5(n_a,n_b) & = &  E(n_as_{\frac{1}{2}}) + E(n_bd_{\frac{5}{2}}) -  E(n_{\text{A}} p_{\frac{3}{2}})-  E(n_{\text{B}} p_{\frac{3}{2}}),\nonumber\\
\delta_6(n_a,n_b) & = &  E(n_ad_{\frac{3}{2}}) + E(n_bd_{\frac{3}{2}}) -  E(n_{\text{A}} p_{\frac{3}{2}})-  E(n_{\text{B}} p_{\frac{3}{2}}),\nonumber\\
\delta_7(n_a,n_b) & = &  E(n_ad_{\frac{3}{2}}) + E(n_bs_{\frac{1}{2}}) -  E(n_{\text{A}} p_{\frac{3}{2}})-  E(n_{\text{B}} p_{\frac{3}{2}}),\nonumber\\
\delta_8(n_a,n_b)  & = & E(n_as_{\frac{1}{2}}) + E(n_bd_{\frac{3}{2}}) -  E(n_{\text{A}} p_{\frac{3}{2}})-  E(n_{\text{B}} p_{\frac{3}{2}}),\nonumber\\
\delta_9(n_a,n_b) & = &  E(n_as_{\frac{1}{2}}) + E(n_bs_{\frac{1}{2}}) -  E(n_{\text{A}} p_{\frac{3}{2}})-  E(n_{\text{B}} p_{\frac{3}{2}}).\nonumber\\
\label{eq101504}
 \end{eqnarray}
Notice that the last three channels, although not relevant to our special initial state with $| r_{\text{A}}r_{\text{B}}\rangle$ because of the conservation of the total angular momenta, are listed for the purpose of completeness and clarity.
 
For the gate by van der Waals interaction, we can set $|r_1\rangle\equiv |80p_{3/2},m_J=3/2\rangle$ and $|r_2\rangle\equiv |90p_{3/2},m_J=3/2\rangle$. With the interaction channels listed above and using the method detailed in Refs.~\cite{Walker2008,Shi2014}, we can calculate the interaction coefficient for $|r_1r_1\rangle$ to be $C_6^{(r_1r_1)}/2\pi=289$~GHz$\mu m^6$. For the state $|r_1r_2\rangle$, there are both a direct energy shift and a state-flip interaction with the other state $|r_2r_1\rangle$. However, the latter effect is slower than the former by four orders of magnitude, thus can be ignored. Hence we only consider the diagonal interaction coefficient for $|r_1r_2\rangle$, which is $C_6^{(r_1r_2)}/2\pi=-281$~GHz$\mu m^6$. The crossover distances~\cite{Saffman2010,Shi2014} for the interaction to be in the van der Waals regime are 0.74~$\mu m$ and $1.3~\mu m$ for $|r_1r_1\rangle$ and $|r_1r_2\rangle$, respectively, which means that a two-qubit separation shall be larger than $1.3\mu m$ to apply the picture of van der Waals interaction.

%=================================================
%end of case 2
%=================================================

\section{Fidelity errors of the gate by van der Waals interaction}\label{App0D}
The fidelity of a quantum logic gate is an important factor that limits whether or not it is useful for a reliable quantum computer~\cite{Preskill1998,Knill2005}. Below we analyze the gate errors of the protocol by van der Waals interaction. There are two sources of error for the gate fidelity, namely, the Rydberg state decay and the population leakage to nearby unwanted transition channels.

\begin{table}
  \begin{tabular}{|c|c|c|}
    \hline   input state  & $T_{\text{Ry}}(r_1)$  & $T_{\text{Ry}}(r_2)$ \\ \hline
  $ |00\rangle$ & 0& 0 \\ \hline 
  $ |01\rangle$ & $\frac{\pi}{\Omega_1} $& $ \frac{\pi}{\Omega_2}$ \\ \hline 
  $ |10\rangle$ &$\frac{\pi}{\Omega_0}+\frac{2\pi}{\Omega_1}+\frac{2\pi}{\Omega_2}$& 0 \\ \hline 
  $ |11\rangle$ & $\frac{\pi}{\Omega_0}+\frac{2\pi}{\Omega_1}+\frac{9}{8}\frac{2\pi}{\Omega_2}$&  $\frac{1}{8}\frac{2\pi}{\Omega_2}$ \\ \hline 
  \end{tabular}
  \caption{ Times for the atom to be in Rydberg states for different input states of the gate using van der Waals interaction. \label{table2}
  }
\end{table}

%\subsection{Gate errors of the protocol by dipolar interaction}
\subsection{Errors due to the decay of Rydberg states}\label{error2edorder01}
The Rydberg state decay induces an error~(probability of de-population) $\tau^{-1} T_{\text{Ry}}$ that is proportional to the time, $T_{\text{Ry}}$, for the state to be in the Rydberg state~\cite{Saffman2005}, as demonstrated by numerical simulation of time evolution of the density matrix in Ref.~\cite{Zhang2012}. Here $\tau$ is the lifetime of the Rydberg state. Because for different input states, $T_{\text{Ry}}$'s are different, we thus
tabulate $T_{\text{Ry}}$'s in Table~\ref{table2} for different input states, where the result $T_{\text{Ry}}(r_1)$ for the input state $|11\rangle$ is partly from numerical calculation: during Pulse-2, the time for $|r_1r_1\rangle$ to be populated is $\frac{1}{8}\frac{2\pi}{\Omega_1}$. Similarly, $T_{\text{Ry}}(r_2)$ for the input state $|11\rangle$ can be found from numerical calculation: during Pulse-3 and Pulse-4, the time for the state $|r_1r_2\rangle$ to be populated is $\frac{1}{8}\frac{2\pi}{\Omega_2}$.  From Table~\ref{table2}, the average probability to have Rydberg state decay of the gate is thus
\begin{eqnarray}
E_{\text{decay}}\approx \frac{\pi}{\tau_1} \left[\frac{1}{2\Omega_0} + \frac{5}{4\Omega_1}+ \frac{17}{16\Omega_2} \right]+\frac{\pi}{\tau_2} \frac{5}{16\Omega_2}, \nonumber
\end{eqnarray}
where $(\tau_1,\tau_2)$ are the lifetimes of the states $|r_1\rangle$ and $|r_2\rangle$, respectively. For the choice of $|r_1\rangle\equiv |80p_{3/2},m_J=3/2\rangle$ and $|r_2\rangle\equiv |90p_{3/2},m_J=3/2\rangle$, we estimate $(\tau_1,\tau_2) = (1.29,1.86)$~ms when $T=4$~K, or $(249,331)~\mu$s when $T=300$~K. Below we assume
\begin{eqnarray}
\Omega_0=\Omega_1.\nonumber
\end{eqnarray} 
Here $\Omega_1= V_1/\sqrt3$, while $V_1=C_6^{(r_1r_1)}/l^6$. Similarly $\Omega_2= V_2/\sqrt3=C_6^{(r_1r_2)}/(\sqrt3l^6)$. As soon as $l$ is set, $\Omega_1$ and $\Omega_2$ are determined,

\begin{figure}
\includegraphics[width=3.4in]
{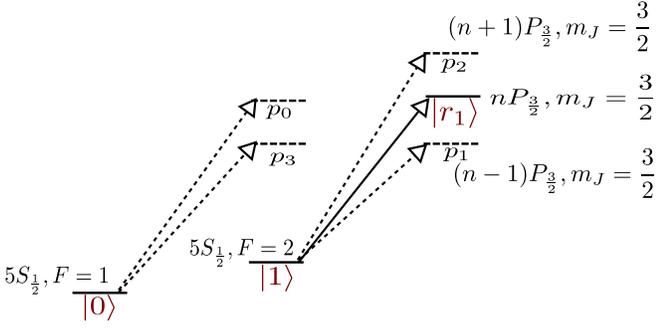}
 \caption{Schematic of the leaking channels for the gate protocol using van der Waals interaction. The leaking channels are denoted by arrowed dashed lines, while the arrowed solid lines denote the excitation of the Rydberg state $|r_1\rangle$. \label{S05} }
\end{figure}

\subsection{Errors due to population leakage to other unwanted transitions}\label{error2edorder02}
There are atomic energy levels quite near to the energy levels used in the gate protocol. Any unwanted transition involving these nearby levels will cause population leakage out of the qubit basis states, as well as phase errors. Their contribution to the gate fidelity error can be analyzed by numerical simulation of the gate protocol. For definiteness, we consider right-hand polarized laser excitation upon the atoms. The two Rydberg states are $|r_1\rangle\equiv |n_1p_{3/2},m_J=3/2,m_I=3/2\rangle$ and $|r_2\rangle\equiv |n_2p_{3/2},m_J=3/2,m_I=3/2\rangle$, respectively, where $(n_1,n_2)=(80,90)$ and the C-six interaction coefficients are given in Appendix~\ref{S-vdWI}.

First of all, although the chosen excitation channel is $|1\rangle\leftrightarrow|r_{1}\rangle$, there is a leaking channel $|0\rangle\leftrightarrow|p_0\rangle$ during Pulse-1 and Pulse-5. Here $|p_0\rangle$ is a superposition state of $|n_1p_{3/2},m_J=1/2,m_I=3/2\rangle$, $|n_1p_{3/2},m_J=3/2,m_I=1/2\rangle$ and $|n_1p_{1/2},m_J=1/2,m_I=3/2\rangle$. We choose Pulse-1 to derive its Rabi frequency, while those for other pulses are similarly derived. A schematic is shown in Fig.~\ref{S05}. The transition $|1\rangle\leftrightarrow|r_1\rangle$ has a Rabi frequency 
\begin{eqnarray}
\Omega_0 &=& e \mathscr{E}(5S ||r|| n_1P) \sum_{m_J}C_{3/2,\overline{q},m_J}^{3/2,1,1/2}C_{m_J,3/2,2}^{1/2,3/2,2}\nonumber\\ &&\times 2 \left\{\begin{array}{ccc}0& 1& 1\\
  3/2&1/2&1/2\end{array} \right\},\label{S07}
\end{eqnarray}
while $|0\rangle\leftrightarrow|p_0\rangle$ has a Rabi frequency 
\begin{eqnarray}
\Omega_0^{(\text{lea})} &=& \sqrt{ (\Omega^{(\text{lea1})})^2 +  (\Omega^{(\text{lea2})})^2+  (\Omega^{(\text{lea3})})^2} ,\label{S08}
\end{eqnarray}
where $\Omega^{(\text{lea1})}$, $\Omega^{(\text{lea2})}$ and $\Omega^{(\text{lea3})}$ are the Rabi frequencies for the couplings of $|n_1p_{3/2},m_J=1/2,m_I=3/2\rangle$, $|n_1p_{3/2},m_J=3/2,m_I=1/2\rangle$ and $|n_1p_{1/2},m_J=1/2,m_I=3/2\rangle$, respectively. 
\begin{eqnarray}
\Omega^{(\text{lea1})} &=& e \mathscr{E}(5S ||r|| n_1P) \sum_{m_J}C_{1/2,\overline{q},m_J}^{3/2,1,1/2}C_{m_J,3/2,1}^{1/2,3/2,1}\nonumber\\ &&\times 2 \left\{\begin{array}{ccc}0& 1& 1\\
3/2&1/2&1/2\end{array} \right\},\nonumber\\
\Omega^{(\text{lea2})} &=& e \mathscr{E}(5S ||r|| n_1P) \sum_{m_J}C_{3/2,\overline{q},m_J}^{3/2,1,1/2}C_{m_J,1/2,1}^{1/2,3/2,1}\nonumber\\ &&\times 2 \left\{\begin{array}{ccc}0& 1& 1\\
3/2&1/2&1/2\end{array} \right\},\nonumber\\
\Omega^{(\text{lea3})} &=& -e \mathscr{E}(5S ||r|| n_1P) \sum_{m_J}C_{1/2,\overline{q},m_J}^{1/2,1,1/2} C_{m_J,3/2,1}^{1/2,3/2,1}\nonumber\\ &&\times \sqrt2 \left\{\begin{array}{ccc}0& 1& 1\\
1/2&1/2&1/2\end{array} \right\}.\label{S09}
\end{eqnarray}
where $q=1,\overline{q}=-1$, $r$ is $(x+iy)/\sqrt2$ here in the presence of right-hand polarized laser field, $(\cdots ||r|| \cdots)$ is a reduced matrix element, $e$ is the elementary charge, $C$ is a Clebsch-Gordan coefficient, $\{\cdots\}$ is a 6-j symbol, and $\mathscr{E}$ is the field strength. From Eqs.~(\ref{S07}),~(\ref{S08}), and~(\ref{S09}), we can have $|\Omega_0^{(\text{lea})}/\Omega_0| =1$.
%\begin{eqnarray} |\Omega_0^{(\text{lea})}/\Omega_0| &=&0.866 . \end{eqnarray}

Because the energy difference between the $n_1p_{3/2}$ state and the $(n_1-1)p_{3/2}$ state is only $\Delta_{p_3}=-15\times2\pi$~GHz, it is necessary to include the leaking channel $|0\rangle\leftrightarrow|p_3\rangle$ during Pulse-1 to Pulse-5. Here $|p_3\rangle$ is a superposition state of $|(n_1-1)p_{3/2},m_J=1/2,m_I=3/2\rangle$, $|(n_1-1)p_{3/2},m_J=3/2,m_I=1/2\rangle$ and $|(n_1-1)p_{1/2},m_J=1/2,m_I=3/2\rangle$, where the superposition obeys a similar relation in the definition of $|p_0\rangle$. Likewise, the Rabi frequency for this leaking channel is
\begin{eqnarray}
\Omega_0^{(\text{lea3})}&=&(1+1/n_1)^{3/2}\Omega_0^{(\text{lea})}.\nonumber
\end{eqnarray}
In Fig.~\ref{S05}, the two levels $|p_3\rangle$ and $|p_1\rangle$ have the same energy, as well as the two levels $|p_0\rangle$ and $|r_1\rangle$.

The two levels $|p_1\rangle$ and $|p_2\rangle$ around $|r_1\rangle$ can also be coupled with $|1\rangle$, with Rabi frequencies $\Omega_{0;r_1}^{(p_1)}$ and $\Omega_{0;r_1}^{(p_2)}$, respectively. Because $(5S ||r|| n_1P)\propto n_1^{-3/2}$, as from Ref.~\cite{Theis2016}, we have 
\begin{eqnarray}
\Omega_{0;r_1}^{(p_1)} = (1+1/n_1)^{3/2}\Omega_0, ~\Omega_{0;r_1}^{(p_2)} = (1+1/n_1)^{-3/2}\Omega_0.\nonumber
\end{eqnarray}

A similar analysis can give us the Rabi frequencies of the leaking channels for Pulse-2 and 3, while those for Pulse-4 and 5 are the same with those of Pulse 3 and 1.

With these leaking channels, the Hamiltonian can be written as
\begin{eqnarray}
  H_{\text{pro-2}} &=& V_1|r_1r_1\rangle\langle r_1r_1| + V_2|r_1r_2\rangle\langle r_1r_2| \nonumber\\
  && + H_{\text{s2}} +H_{00} +H_{01} +H_{10} +H_{11}\nonumber\\
  && -2\omega_{\text{g}}|00\rangle\langle 00|-\omega_{\text{g}}\sum_{\varphi}(|0\varphi\rangle\langle 0\varphi|+|\varphi0\rangle\langle \varphi0|)\nonumber\\
  && -(\omega_{\text{g}}-\Delta_{\varsigma})\sum_{\varsigma}(|0\varsigma\rangle\langle 0\varsigma|+|\varsigma0\rangle\langle \varsigma0|), \label{Hpro2}
\end{eqnarray}
where $\varphi=1,p_0,p_3,p_0',p_3'$, and $\varsigma=p_1,p_2,p_1',p_2'$. Here $|p_0'\rangle$ and $|p_3'\rangle$ are states similar to $|p_0\rangle$ and $|p_3\rangle$, with $n_1$ replaced by $n_2$ in their respective definitions, and $|p_1'\rangle$ and $|p_2'\rangle$ are the two states below and above $|r_2\rangle$ by a difference of one in their principal quantum numbers. Here $H_{\text{s2}}$ is the Hamiltonian for the gate sequence,
\begin{eqnarray}
  H_{\text{s2}}: \left\{\begin{array}{ll}
\Omega_0(|10\rangle\langle r_10|+ |11\rangle\langle r_11|)/2 +\text{h.c.},& \text{Pulse-1;}\\
\Omega_1(|01\rangle\langle 0r_1|+ |r_11\rangle\langle r_1r_1|)/2 +\text{h.c.},& \text{Pulse-2;}\\
\Omega_2(|01\rangle\langle 0r_2|+ |r_11\rangle\langle r_1r_2|)/2 +\text{h.c.},& \text{Pulse-3;}\\
-\Omega_2(|01\rangle\langle 0r_2|+ |r_11\rangle\langle r_1r_2|)/2 +\text{h.c.},& \text{Pulse-4;}\\
\Omega_0(|10\rangle\langle r_10|+ |11\rangle\langle r_11|)/2 +\text{h.c.},& \text{Pulse-5,}
  \end{array}\right. \nonumber
\end{eqnarray} 
and $\{H_{00},H_{01},H_{10},H_{11}\}$ account for the leakage of the population. $H_{00},H_{01},H_{10}$ and $H_{11}$ are respectively given by
%\onecolumngrid
%\begin{widetext}
\begin{eqnarray}
&& \left\{\begin{array}{ll}
|00\rangle[\Omega_0^{(\text{lea})} \langle p_00| +\Omega_0^{(\text{lea3})} \langle p_30|]/2 +\text{h.c.},& \text{Pulse-1;}\\
|00\rangle[\Omega_1^{(\text{lea})} \langle 0p_0|+\Omega_1^{(\text{lea3})} \langle 0p_3|]/2 +\text{h.c.},& \text{Pulse-2;}\\
|00\rangle[\Omega_2^{(\text{lea})} \langle 0p_0'|+\Omega_2^{(\text{lea3})} \langle 0p_3'|]/2 +\text{h.c.},& \text{Pulse-3;}\\
-|00\rangle[\Omega_2^{(\text{lea})}\langle 0p_0'|+\Omega_2^{(\text{lea3})} \langle 0p_3'|]/2 +\text{h.c.},& \text{Pulse-4;}\\
|00\rangle[\Omega_0^{(\text{lea})} \langle p_00| +\Omega_0^{(\text{lea3})} \langle p_30|]/2 +\text{h.c.},& \text{Pulse-5,}
  \end{array}\right. \nonumber\\
&&   \left\{\begin{array}{ll}
 |01\rangle   [\Omega_0^{(\text{lea})} \langle p_01| +\Omega_0^{(\text{lea3})} \langle p_31|]/2/2 +\text{h.c.},& \text{Pulse-1;}\\
\sum_j\Omega_{1;r_1}^{(p_j)} |01\rangle\langle 0p_j|/2 +\text{h.c.},& \text{Pulse-2;}\\
\sum_j\Omega_{2;r_2}^{(p_j')} |01\rangle\langle 0p_j'|/2 +\text{h.c.},& \text{Pulse-3;}\\
-\sum_j\Omega_{2;r_2}^{(p_j')} |01\rangle\langle 0p_j'|/2 +\text{h.c.},& \text{Pulse-4;}\\
|01\rangle   [\Omega_0^{(\text{lea})} \langle p_01| +\Omega_0^{(\text{lea3})} \langle p_31|]/2/2 +\text{h.c.} ,& \text{Pulse-5,}
  \end{array}\right. \nonumber\\
&&   \left\{\begin{array}{ll}
\sum_j\Omega_{0;r_1}^{(p_j)} |10\rangle\langle p_j 0|/2 +\text{h.c.},& \text{Pulse-1;}\\
|r_10\rangle [\Omega_1^{(\text{lea})} \langle r_1p_0| +\Omega_1^{(\text{lea3})} \langle r_1p_3|  ]/2 +\text{h.c.},& \text{Pulse-2;}\\
|r_10\rangle[\Omega_2^{(\text{lea})} \langle r_1p_0'| + \Omega_2^{(\text{lea3})} \langle r_1p_3'| ]/2 +\text{h.c.},& \text{Pulse-3;}\\
-|r_10\rangle[\Omega_2^{(\text{lea})} \langle r_1p_0'| + \Omega_2^{(\text{lea3})} \langle r_1p_3'| ]/2  +\text{h.c.},& \text{Pulse-4;}\\
\sum_j\Omega_{0;r_1}^{(p_j)} |10\rangle\langle p_j 0|/2 +\text{h.c.},& \text{Pulse-5,}
  \end{array}\right. \nonumber\\
&& \left\{\begin{array}{ll}
\sum_j\Omega_{0;r_1}^{(p_j)} |11\rangle\langle p_j 1|/2 +\text{h.c.},& \text{Pulse-1;}\\
\sum_j\Omega_{1;r_1}^{(p_j)} |r_11\rangle\langle r_1p_j|/2 +\text{h.c.},& \text{Pulse-2;}\\
\sum_j\Omega_{2;r_2}^{(p_j')} |r_11\rangle\langle r_1p_j'|/2 +\text{h.c.},& \text{Pulse-3;}\\
-\sum_j\Omega_{2;r_2}^{(p_j')} |r_11\rangle\langle r_1p_j'|/2 +\text{h.c.},& \text{Pulse-4;}\\
\sum_j\Omega_{0;r_1}^{(p_j)} |11\rangle\langle p_j 1|/2 +\text{h.c.},& \text{Pulse-5.}
  \end{array}\right. \nonumber
\end{eqnarray} 
In the analysis above, we have neglected the van der Waals interaction of the states like $|r_1p_k\rangle$ and $|r_1p_k'\rangle$, where $k=0,1$ or $2$. The reason is that their van der Waals interaction is small compared with the detunings of the leaking transitions. Thus inclusion of the van der Waals interaction of the states like $|r_1p_k\rangle$ and $|r_1p_k'\rangle$ will not alter our conclusion. Moreover, we have imposed a nonzero energy for the level $|0\rangle$ in order to eliminate the time dependence in the Hamiltonian of the leaking channels in a rotating frame. Now $|0\rangle$ has an energy of $-\omega_{\text{g}}$, thus the time evolution during the gate sequence will add a phase term $e^{2i\omega_{\text{g}} t_{\text{g}}}$ to the input state $|00\rangle$, and $e^{i\omega_{\text{g}} t_{\text{g}}}$ to the input states $|01\rangle$ and $|10\rangle$, where $t_{\text{g}}$ is the gate operation time. In this case, the gate will transform the input states as $\{|00\rangle,|01\rangle,|10\rangle,|11\rangle\}\mapsto \{e^{2i\omega_{\text{g}} t_{\text{g}}}|00\rangle,-e^{i\omega_{\text{g}} t_{\text{g}}}|01\rangle,-e^{i\omega_{\text{g}} t_{\text{g}}}|10\rangle,-|11\rangle\}$, which will be used in the numerical calculation of the gate fidelity.

Another possible error may arise due to the atomic heating in the presence of Rydberg-Rydberg interaction. In Fig.~\ref{S02}, we notice that the state $|r_1r_1\rangle$ can be populated during Pulse-2. This means that there will be a repulsion~(during Pulse-2; attraction during Pulse-3 and 4) between the two atoms due to the van der Waals interaction,
\begin{eqnarray}
 F(l)&=& \frac{dV_1(l)}{dl}=-6C_6(l)/l^7,\nonumber
\end{eqnarray} 
where $C_6=289\times2\pi$~GHz$\mu m^6$, and $l$ is the two-atom separation. The relative speed of the two atoms will change by 
\begin{eqnarray}
 \delta v&=& 6C_6(l) T_{r_1r_1}/(\mu l^7),\label{deltaV}
\end{eqnarray} 
where $\mu=87u$ is the mass of the atom, $u$ is the atomic mass unit, and $T_{r_1r_1}=\pi/(4\Omega_1)$ is the total time for the state $|r_1r_1\rangle$ to be populated during Pulse-2, as numerically calculated. For a typical parameter setting $l=4.36\mu m$ and $\Omega_1/2\pi=24.23$~MHz, we can estimate $\delta v\approx 0.23$~nm$/\mu s$. With a typical total gate time of about $t_{\text{g}}=0.1\mu s$, we have a change of the two-atom separation $\lesssim \delta vt_{\text{g}} /2\approx0.01$~nm, if the initial two-atom relative velocity is zero. This change of the two-atom separation is orders of magnitude smaller than $l$, thus can be ignored. Similar conclusion can be drawn about the Rydberg-interaction-induced atomic motion in the other pulses of the gate sequence.

We denote the gate transformation by $\mathscr{G}:\{|00\rangle,|01\rangle,|10\rangle,|11\rangle\}\mapsto \{|00\rangle,-|01\rangle,-|10\rangle,-|11\rangle\}$. The error of the quantum gate can be conveniently defined as an average of errors for each transformation of the four basis states. Numerically, we can use the Hamiltonian $H_{\text{pro-2}}$ upon the four different input states for time evolution simulation, and define their respective errors as
\begin{eqnarray}
E_{s}&=& 1-\left|\langle s| \mathscr{G}^\dag\int dt e^{-itH_{\text{pro-2}}(t)}|s\rangle\right|^2,\label{Es}
\end{eqnarray}
where $s\in\{00,01,10,11\}$. Due to leakage out of the transition channels, there will be a leakage error for the gate protocol 
\begin{eqnarray}
  E_{\text{leak}}&=& \sum_s E_{s}/4.\label{Eleak}
\end{eqnarray}

In conclusion, the total gate error is a sum of the errors due to Rydberg state decay and population leakage,
\begin{eqnarray}
 E&=&E_{\text{decay}} +  E_{\text{leak}}.\nonumber
\end{eqnarray}
We have conducted numerical simulation for the gate fidelity error, where the results are presented in Fig.~\ref{fig03}. The inset of Fig.~\ref{fig03} shows that it is possible to have a gate fidelity error of $5.28\times10^{-5}$ with using $\Omega/2\pi=24.23$~MHz. Importantly, the two strengths of the van der Waals interaction here are only $(V_1,V_2)=(41.97,-41.81)\times2\pi$~MHz. To have these interaction strengths, the corresponding two-atom separation is $l=4.36~\mu m$. Note that as discussed below Eq.~(\ref{eq101504}), the crossover distance for the two atom interaction to be in the van der Waals regime is $1.3\mu m$. This means that a two-atom separation of $4.36\mu m$ of course satisfies the condition of using van der Waals interaction.

  \section{Fluctuation of atomic positions}\label{fluctuationError}
All the analysis here considers ideal experimental conditions. For the current technology of optical dipole traps, the finite depths of the traps for trapping the neutral atoms will also give some extra error.  This is because the blockade shift $V_1$ and $V_2$ depends on the two-atom separation $l$. Here we will analyze the error due to this effect, where the atomic spatial distribution is determined by the trap parameters and the atomic motional states.

\begin{figure}
\includegraphics[width=2.1in]
{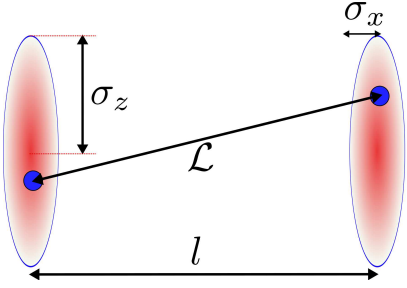}
\caption{Schematic of position fluctuation of atoms. An atom is mainly trapped inside the ellipse. $\sigma_j^2$ is the variance of the atomic position along the $j$-axis, where $j=x,y$ or $z$. Here $\sigma_x=\sigma_y$. \label{fig-position01}}
\end{figure}

\begin{figure}
\includegraphics[width=3.3in]%, angle=-90]
{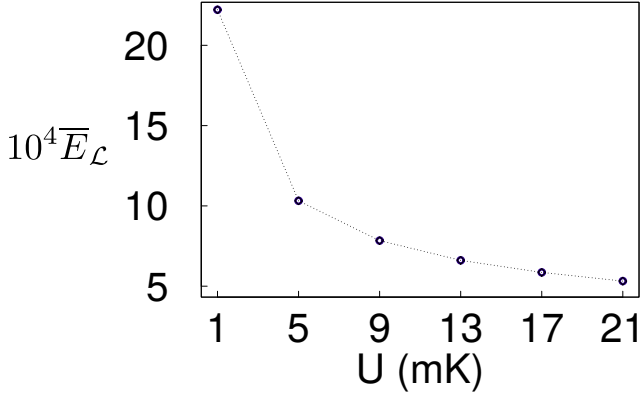}
\caption{Scaled gate fidelity error due to the fluctuation of atomic positions as a function of the trap depth when $l=4.36\mu m$.  \label{p2}}
\end{figure}

A commonly used method of trapping a single neutral atom in Rydberg experiments is far-off-resonance optical trap~\cite{Isenhower2010}, or optical tweezer~\cite{Wilk2010}. With optical tweezers, the authors of Ref.~\cite{Kaufman2012} successfully employed Raman sideband cooling to cool an $^{87}$Rb atom to its motional ground state. The fact that people can cool neutral atoms to their motional ground states~\cite{Kaufman2012} and load neutral atoms efficiently to optical tweezer lattice of a small lattice constant~\cite{Lester2015} is the background of the discussion in this appendix. The parameters characterizing an optical trap include the trap depth $U$ and the $1/e^2$ beam radius $w$, which further determines the qubit's oscillation frequencies $\{\omega_x,\omega_y,\omega_z\}$, and the averaged variances, $\{\sigma_x^2,\sigma_y^2,\sigma_z^2\}$, of its position.

When the motional state of a trapped neutral atom is thermal, i.e., $k_BT/2\geq \hbar \omega_j$, $j=x,y,z$, the position fluctuation of the trapped atom can be as large as several microns~\cite{Isenhower2010}. Because the atomic separation $l$ is only several micrometers in our gate, we conclude that in this regime the gate protocol will have large fidelity error due to its position fluctuation.

When the motional state of a trapped neutral atom is non-classical, i.e., $k_BT/2\leq \hbar \omega_j$, $j=x,y,z$, a trapped atom was ever cooled to its motional ground state characterized with zero vibration excitation~\cite{Kaufman2012}. When the atom is in a state with $\mathcal{N}$ vibration excitations, its position variances are
\begin{eqnarray}
\sigma_j^2 &=&\left(\mathcal{N}+\frac{1}{2}\right) \frac{\hbar}{\mu\omega_j}.\nonumber
\end{eqnarray}%where $\mu$ is the mass of the atom, which is equal to $87u$, with $u$ the atomic mass unit, for an $^{87}$Rb atom.
For a typical $\omega_j=150\times 2\pi$~kHz, we have $\sigma_j=19.6$~nm when $\mathcal{N}=0$. Consider an optical tweezer studied in Ref.~\cite{Kaufman2012} for trapping and cooling an $^{87}$Rb atom, the trap frequencies along the $x$ and $y$ directions are theoretically given by
\begin{eqnarray}
\omega_x\approx\omega_y = \frac{2}{w}\sqrt{U/\mu},\nonumber
\end{eqnarray}
which gives $\omega_x\approx\omega_y=2\pi\times153$~kHz if we use $U=1.4$mK and $w=0.76\mu$m for the example of Ref.~\cite{Kaufman2012}. The measured result in Ref.~\cite{Kaufman2012} is $\{\omega_z,\omega_x,\omega_y\}=$$\{30,154,150\}$~kHz, which means that $\omega_z\approx \omega_x/5$. Concerning whether it is possible to trap two neutral atoms as close as the $l$ used in this work, we note that  the authors in Ref.~\cite{Lester2015} created arrays of deeper optical tweezers, where each trap is characterized by $U=73$~MHz$\sim3.5$~mK and $w=0.71\mu$m, and successfully loaded $^{87}$Rb atoms in small arrays with optical lattice constant as small as $1.7\mu$m.

As shown in Fig.~\ref{fig-position01}, the actual distance $\mathcal{L}$ between the two atoms can be different from the wanted distance $l$. To describe this, we denote the wanted position of the control and target qubits as $(0,0,0)$ and $(l,0,0)$, respectively. When the motional state of the control qubit inside an optical tweezer is the ground state, the distribution of its actual location is,
\begin{eqnarray}
  f_c(x_c,y_c,z_c) &=& \text{exp} \left[-0.5(x_c^2+y_c^2)/\sigma_x^2 -0.5z_c^2/\sigma_z^2\right]\nonumber\\
  &&/\left[\sigma_x^2\sigma_z(2\pi)^{3/2}\right],\nonumber
\end{eqnarray}
while that for the target qubit is
\begin{eqnarray}
  f_t(x_t,y_t,z_t) &=& \text{exp} \left[-0.5((x_t-l)^2+y_t^2)/\sigma_x^2 -0.5z_t^2/\sigma_z^2\right]\nonumber\\
  &&/\left[\sigma_x^2\sigma_z(2\pi)^{3/2}\right].\nonumber
\end{eqnarray}
Here the subscript c~(t) denotes control~(target).

For different runs of the gate cycles, the fluctuation of the atomic location will give different $\mathcal{L}$'s and different orientations of the two-atom axis relative to the quantization axis $x$, where $\mathcal{L}^2=(x_c-x_t)^2+(y_c-y_t)^2+(z_c-z_t)^2$. For $\sigma_z\ll l$, we mainly focus on $\mathcal{L}$'s fluctuation and its consequence upon the gate performance.  For a $C_Z$ gate, the deviation of the blockade shift from $V_1(l)$ and $V_2(l)$, i.e., the blockade shift of the state $|r_1r_1\rangle$ and $|r_1r_2\rangle$, will contribute an extra error $E_{\mathcal{L}}(\mathcal{L})$ to the total gate fidelity error, which can be numerically evaluated. The average of  $E_{\mathcal{L}}(\mathcal{L})$ is
\begin{eqnarray}
  \overline{E}_{\mathcal{L}}&=&\int dx_c\int dy_c \cdots \int dy_t\int dz_t E_{\mathcal{L}}(\mathcal{L})f_c f_t .  \nonumber
\end{eqnarray}
The above integration as a function of $U$ can be performed by Monte Carlo integration, where a test can be made by checking if $\overline{E}_{\mathcal{L}}$ becomes unit when we set $E_{\mathcal{L}}(\mathcal{L})=1$ in the integral above. The numerical result of $\overline{E}_{\mathcal{L}}$ is presented in Fig.~\ref{p2} for several values of $U$, with the parameters $\Omega/2\pi=24.23$~MHz, and the two van der Waals interactions $(V_1,V_2)=(41.97,-41.81)\times2\pi$~MHz~[corresponding to $l=4.36\mu m$]. One can find that $\overline{E}_{\mathcal{L}}$ drops from $2.2\times10^{-3}$ to $5.3\times10^{-4}$ when $U$ increases from $1$~mK to $21$~mK. In these cases, $\overline{E}_{\mathcal{L}}$ is much larger than the intrinsic gate fidelity error of about $5\times10^{-5}$ presented in Fig.~\ref{fig03}, thus becomes the major error of the gate fidelity. Notice that the trap depth $U$ of a few times of $10$~mK is feasible for the current optical trap technology~\cite{Saffman2016}. Thus it is possible to realize a trap with $U\sim20$~mK, so that the total gate fidelity error becomes $1-\mathcal{F}\approx6\times10^{-4}$ with our protocol for an immediate implementation.

%=======================================
%error case 1
%=======================================
\section{Fidelity errors of the gate by dipolar interaction}\label{error1order}
Below we analyze the gate errors of the dipolar-interaction based protocol, where the analysis is similar to that for the gate protocol by van der Waals interaction.
 
\subsection{Errors due to the decay of Rydberg states}
We tabulate $T_{\text{Ry}}$'s in Table~\ref{table1}, where the result for the input state $|11\rangle$ is from numerical calculation: during Pulse-2, the times for the states $|d1\rangle,|pf\rangle,|fp\rangle$, and $|dd\rangle$ to be populated are approximately $(0.594,0.141,0.141,0.125)\frac{2\pi}{\Omega}$, respectively. From Table~\ref{table1}, the average probability to have Rydberg state decay of the gate is thus
\begin{eqnarray}
E_{\text{decay}}\approx \frac{\pi}{\tau_d} \left[\frac{1}{2\Omega_0} + \frac{1.17}{\Omega} \right]+\frac{\pi}{\tau_p} \frac{0.14}{\Omega}+\frac{\pi}{\tau_f} \frac{0.14}{\Omega},\nonumber
\end{eqnarray}
where $(\tau_p,\tau_d,\tau_f)$ are the lifetimes of the three states $|d\rangle,|f\rangle$ and $|p\rangle$. For $n=59$ and an atomic temperature of $T=4$~K, one can use the result of Ref.~\cite{Saffman2005} to estimate $(\tau_p,\tau_d,\tau_f) = (557,196,97)\mu s$. Below we assume $\Omega_0=\Omega$.

\begin{table}
  \begin{tabular}{|c|c|c|c|}
    \hline   input state  & $T_{\text{Ry}}(d)$  & $T_{\text{Ry}}(p)$  & $T_{\text{Ry}}(f)$   \\ \hline
  $ |00\rangle$ & 0& 0& 0 \\ \hline 
  $ |01\rangle$ & $\frac{\pi}{\Omega}$& 0& 0 \\ \hline 
  $ |10\rangle$ &$\frac{\pi}{\Omega_0}+\frac{2\pi}{\Omega}$& 0& 0 \\ \hline 
  $ |11\rangle$ & $\frac{\pi}{\Omega_0}+0.84\frac{2\pi}{\Omega}$& $0.28\frac{2\pi}{\Omega}$& $0.28\frac{2\pi}{\Omega}$ \\ \hline 
  \end{tabular}
  \caption{ Times for the atom to be in Rydberg states for different input states of the gate using dipole-dipole interaction. \label{table1}
  }
\end{table}

\subsection{Errors due to population leakage to other unwanted transitions}
\begin{figure}
\includegraphics[width=3.4in]
{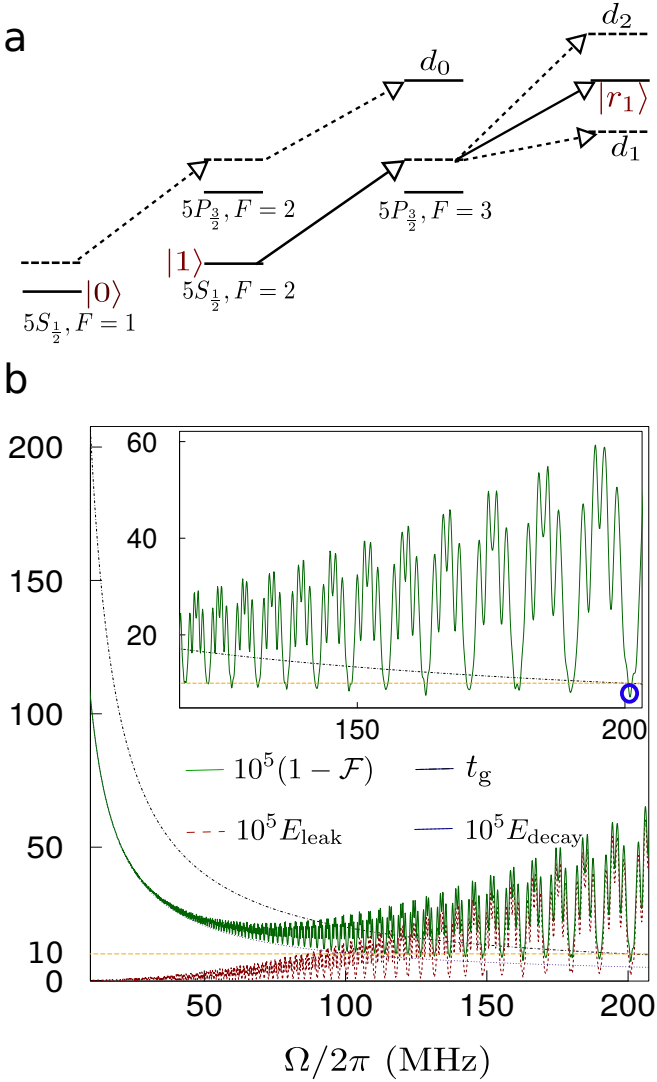}
 \caption{(a) Schematic of the leaking channels for the gate protocol using dipolar interaction. The leaking channels are denoted by arrowed dashed lines, while the arrowed solid lines denote the two-photon transition for the excitation of the Rydberg state $|r_1\rangle$. (b) Performance quality of the gate by using dipolar interaction for different Rabi frequencies $\Omega$'s. Plotted are decay error $E_{\text{decay}}$, leakage error $E_{\text{leak}}$ and the total fidelity error $1-\mathcal{F}$, scaled by $10^5$. The lifetimes of the Rydberg states are estimated with $T=4$~K. We also plot the gate time $t_{\text{g}}=4\pi/\Omega$, with $\Omega_0=\Omega$. Inset: $t_{\text{g}}$ and the scaled fidelity error when $1-\mathcal{F}$ can be smaller than $10^{-4}$, a value denoted by the horizontal dashed line. The circle at the right bottom of the inset denotes a case with $\Omega/2\pi=200.9$~MHz, $t_{\text{g}}=9.96$~ns, and $1-\mathcal{F}=7.16\times10^{-5}$.   \label{S-02} }
\end{figure}

Similar to the gate protocol by using van der Waals interaction, the leakage to nearby levels during the gate sequence will induce error to the gate transformation. For definiteness, we choose $n=59$ and the intermediate state $|e\rangle\equiv |5P_{3/2}, F=3,m_f=3\rangle$ for the excitation of the Rydberg state. The level diagram is shown in Fig.~\ref{S-02}(a). Although the laser excitation is resonant with the transition $|1\rangle\leftrightarrow|r_1\rangle$, the level $|0\rangle$, which is below $|1\rangle$ by $\omega_{\text{g}}/2\pi=6.8$~GHz, will be coupled off-resonantly. For a two-photon transition from the ground state $|1\rangle$ to the Rydberg state $|r_1\rangle$, the Rabi frequency is given by 
\begin{eqnarray}
\Omega_0 \approx \frac{\Omega_{\text{low}}\Omega_{\text{upp}} }{2\delta_{\text{2-pho}}},\nonumber
\end{eqnarray}
where $\Omega_{\text{low}}$ and $\Omega_{\text{upp}}$ are the Rabi frequencies of the lower transition $|1\rangle\leftrightarrow|e\rangle$ and upper transition $|e\rangle\leftrightarrow|r_1\rangle$, respectively, and $\delta_{\text{2-pho}}$ is the detuning for the lower transition. According to Wigner-Eckart theorem, 
\begin{eqnarray}
  \Omega_{\text{low}} &=&-e \mathscr{E}_{\text{low}}(5P_{3/2},F=3 ||r|| 5S_{1/2},F=2) C_{2,q,3}^{2,1,3} \nonumber\\ &=&-e \mathscr{E}_{\text{low}}(5P_{3/2} ||r|| 5S_{1/2}) C_{2,q,3}^{2,1,3} \nonumber\\ &&\times 2\sqrt5 \left\{\begin{array}{ccc}3/2& 1/2& 1\\
  2&3&3/2\end{array} \right\} , \nonumber\\
\Omega_{\text{upp}} &=&-e \mathscr{E}_{\text{upp}}(5P_{3/2} ||r|| nD_{5/2}) \sum_{m_J}C_{5/2,\overline{q},m_J}^{5/2,1,3/2}C_{m_J,3/2,3}^{3/2,3/2,3}\nonumber\\ &=&e \mathscr{E}_{\text{upp}}(5P ||r|| nD) \sum_{m_J}C_{5/2,\overline{q},m_J}^{5/2,1,3/2}C_{m_J,3/2,3}^{3/2,3/2,3}\nonumber\\ &&\times 3\sqrt2 \left\{\begin{array}{ccc}1& 2& 1\\
  5/2&3/2&1/2\end{array} \right\},\nonumber
\end{eqnarray}
where $m_J\in\{-3/2,\cdots,3/2\}$ for the intermediate state. For a $\delta_{\text{2-pho}}$ of several GHz, the unwanted transition $|0\rangle\leftrightarrow|r_1\rangle$ will have a Rabi frequency $\Omega_{0}^{(\text{lea})}\approx \frac{\Omega_{\text{low}}'\Omega_{\text{upp}}' }{2\delta_{\text{2-pho}}}$. It seems that we will have an unwanted transition $|0\rangle\leftrightarrow|5P_{3/2}, F=3,m_f=2\rangle$ with Rabi frequency $\Omega_{\text{low}}'=-e \mathscr{E}(5P_{3/2} ||r|| 5S_{1/2}) C_{1,q,2}^{1,1,3}/\hbar$. Nevertheless, we have $C_{1,1,2}^{1,1,3}=0$, which means that there will be no coupling between $|0\rangle$ and the chosen intermediate level. Now the hyperfine splitting between $F=2$ and $F=1$ levels of the state $5P_{3/2}$ is about $267\times2\pi$~MHz, which means that there can be some coupling between $|0\rangle$ and $|5P_{3/2}, F=2,m_f=2\rangle$, which can be further coupled with $|d_0\rangle=(\zeta_1|nD_{5/2},m_J=3/2,m_I=3/2\rangle+\zeta_2|nD_{5/2},m_J=5/2,m_I=1/2\rangle+\zeta_3|nD_{3/2},m_J=3/2,m_I=3/2\rangle)$, a state different from the target Rydberg state $|r_1\rangle=|59D_{3/2},m_J=5/2,m_I=3/2\rangle$, with $\zeta_1,\zeta_2$ and $\zeta_3$ determined by the coupling strengths. Here we include the manifold $D_{3/2}$ because the fine structure splitting for the d-orbital state is smaller than $60\times2\pi$~MHz when $n=59$, much smaller than $\omega_g$. In this case, we have 
\begin{eqnarray}
\Omega_{\text{low}}' &=&-e \mathscr{E}_{\text{low}}(5P_{3/2},F=2 ||r|| 5S_{1/2},F=1) C_{1,q,2}^{1,1,2}\nonumber\\&=&e \mathscr{E}_{\text{low}}(5P_{3/2} ||r|| 5S_{1/2}) C_{1,q,2}^{1,1,2}\nonumber\\&&\times 2\sqrt3 \left\{\begin{array}{ccc}3/2& 1/2& 1\\
  1&2&3/2\end{array} \right\}, \nonumber\\
\Omega_{\text{upp}}' &=&\sqrt{\Omega_{\text{upp1}}^2+ \Omega_{\text{upp2}}^2+ \Omega_{\text{upp3}}^2}, \nonumber
\end{eqnarray}
where $\Omega_{\text{upp1}}$, $\Omega_{\text{upp2}}$ and $\Omega_{\text{upp3}}$ are the Rabi frequencies for the coupling of  $|nD_{5/2},m_J=3/2,m_I=3/2\rangle $, $|nD_{5/2},m_J=5/2,m_I=1/2\rangle $ and $|nD_{3/2},m_J=3/2,m_I=3/2\rangle$, respectively,
\begin{eqnarray}
\Omega_{\text{upp1}} &=& e \mathscr{E}_{\text{upp}}(5P ||r|| nD) \sum_{m_J}C_{3/2,\overline{q},m_J}^{5/2,1,3/2}C_{m_J,3/2,2}^{3/2,3/2,2}\nonumber\\ &&\times 3\sqrt2 \left\{\begin{array}{ccc}1& 2& 1\\
  5/2&3/2&1/2\end{array} \right\},\nonumber\\
\Omega_{\text{upp2}} &=& e \mathscr{E}_{\text{upp}}(5P ||r|| nD) \sum_{m_J}C_{5/2,\overline{q},m_J}^{5/2,1,3/2}C_{m_J,1/2,2}^{3/2,3/2,2}\nonumber\\ &&\times 3\sqrt2 \left\{\begin{array}{ccc}1& 2& 1\\
  5/2&3/2&1/2\end{array} \right\},\nonumber\\
\Omega_{\text{upp3}} &=& -e \mathscr{E}_{\text{upp}}(5P ||r|| nD) \sum_{m_J}C_{3/2,\overline{q},m_J}^{3/2,1,3/2}C_{m_J,3/2,2}^{3/2,3/2,2}\nonumber\\ &&\times 2\sqrt3 \left\{\begin{array}{ccc}1& 2& 1\\
3/2&3/2&1/2\end{array} \right\}.\nonumber
\end{eqnarray}
With these analysis, we can determine the Rabi frequency for the leaking channel,
\begin{eqnarray}
|\Omega_{0}^{(\text{lea})}/\Omega_{0}| &=&| \frac{\Omega_{\text{low}}'\Omega_{\text{upp}}' }{  \Omega_{\text{low}}\Omega_{\text{upp}}}|\approx 0.652.\nonumber
\end{eqnarray}

The analysis above applies for Pulse-1 and Pulse-3. For Pulse-2, a similar analysis can give us the Rabi frequency $\Omega_{1}^{(\text{lea})}$ for the leaking channel $|0\rangle\leftrightarrow|d_0\rangle$.

Population can also leak to Rydberg levels that are near $|r_1\rangle$, contributing errors to the gate fidelity. There are two nearby levels around $|r_1\rangle$: $|d_1\rangle$ and $|d_2\rangle$, which are below and above $|r_1\rangle$ by a difference of principal quantum number $-1$ and $1$, respectively, and identified with detunings $\Delta_{d_1}$ and $\Delta_{d_2}$, respectively.  During Pulse-1 and Pulse-3, the Rabi frequency for the leaking channel $|1\rangle\leftrightarrow|d_1\rangle$ is given by $\Omega_{0;r_1}^{(d_1)}\approx \frac{\Omega_{\text{low}}\Omega_{\text{upp}}' }{2\delta_{\text{2-pho}}}$. To estimate it, we can compare the following rates,
\begin{eqnarray}
\Omega_{\text{upp}} &=&-e \mathscr{E}(5P_{3/2} ||r||  nS_{1/2}) C_{r},\nonumber\\
\Omega_{\text{upp}}' &=&-e \mathscr{E}(5P_{3/2} ||r||(n+1)S_{1/2} ) C_{r},\nonumber
\end{eqnarray} 
where $C_r$, a factor arose from the angular momentum selection rules, is the same for $\Omega_{\text{upp}}$ and $\Omega_{\text{upp}}'$. Now $(5P_{3/2} ||r||  nS_{1/2})\propto n^{-3/2}$~\cite{Saffman2010}, thus we know that $\Omega_{0;r_1}^{(d_1)}/\Omega_{0}=\Omega_{\text{upp}}'/\Omega_{\text{upp}}\approx [1+1/n)]^{3/2}$. Similarly, we have $\Omega_{0}/\Omega_{0;r_1}^{(d_2)}\approx [1+1/n)]^{3/2}$. During Pulse-2, similar leaking occurs for these channels, with Rabi frequencies $\Omega_{1;r_1}^{(d_1)}=[1+1/n)]^{3/2}\Omega_1$ and $\Omega_{1;r_1}^{(d_2)}=[1+1/n)]^{-3/2}\Omega_1$, respectively.

With these leaking channels, the Hamiltonian can be written as
\begin{eqnarray}
  H_{\text{pro-1}} &=& H_{\text{dd}} + H_{\text{s1}} +H_{00} +H_{01} +H_{10} +H_{11}\nonumber\\
  && -2\omega_{\text{g}}|00\rangle\langle 00|-\omega_{\text{g}}\sum_{\varphi}(|0\varphi\rangle\langle 0\varphi|+|\varphi0\rangle\langle \varphi0|)\nonumber\\
  && -(\omega_{\text{g}}-\Delta_{\varsigma})\sum_{\varsigma}(|0\varsigma\rangle\langle 0\varsigma|+|\varsigma0\rangle\langle \varsigma0|),\label{Hpro1}
\end{eqnarray}
where $\varphi=1,d_0$ and $\varsigma=d_1,d_2$, $H_{\text{dd}}$ is defined in Eq.~(\ref{Hdddefine}), $H_{\text{s1}}$ is the Hamiltonian for the gate sequence,
\begin{eqnarray}
  H_{\text{s1}} &=& \left\{\begin{array}{ll}
\Omega_0(|10\rangle\langle r_10|+ |11\rangle\langle r_11|)/2 +\text{h.c.},& \text{Pulse-1;}\\
\Omega_1(|01\rangle\langle 0r_1|+ |r_11\rangle\langle r_1r_1|)/2 +\text{h.c.},& \text{Pulse-2;}\\
\Omega_0(|10\rangle\langle r_10|+ |11\rangle\langle r_11|)/2 +\text{h.c.},& \text{Pulse-3,}
  \end{array}\right. \nonumber
\end{eqnarray} 
and $\{H_{00},H_{01},H_{10},H_{11}\}$ account for the leakage of the population,
\begin{eqnarray}
  H_{00} &=& \left\{\begin{array}{ll}
\Omega_0^{(\text{lea})} |00\rangle\langle d_00|/2 +\text{h.c.},& \text{Pulse-1;}\\
\Omega_1^{(\text{lea})} |00\rangle\langle 0d_0|/2 +\text{h.c.},& \text{Pulse-2;}\\
\Omega_0^{(\text{lea})} |00\rangle\langle d_00|/2 +\text{h.c.},& \text{Pulse-3,}
  \end{array}\right. \nonumber
\end{eqnarray} 
\begin{eqnarray}
  H_{01} &=& \left\{\begin{array}{ll}
\Omega_0^{(\text{lea})} |01\rangle\langle d_01|/2 +\text{h.c.},& \text{Pulse-1;}\\
\sum_j\Omega_{1;r_1}^{(d_j)} |01\rangle\langle 0d_j|/2 +\text{h.c.},& \text{Pulse-2;}\\
\Omega_0^{(\text{lea})} |01\rangle\langle d_01|/2 +\text{h.c.},& \text{Pulse-3,}
  \end{array}\right. \nonumber
\end{eqnarray}
\begin{eqnarray}
  H_{10} &=& \left\{\begin{array}{ll}
\sum_j\Omega_{0;r_1}^{(d_j)} |10\rangle\langle d_j 0|/2 +\text{h.c.},& \text{Pulse-1;}\\
\Omega_1^{(\text{lea})} |r_10\rangle\langle r_1d_0|/2 +\text{h.c.},& \text{Pulse-2;}\\
\sum_j\Omega_{0;r_1}^{(d_j)} |10\rangle\langle d_j 0|/2 +\text{h.c.},& \text{Pulse-3,}
  \end{array}\right. \nonumber
\end{eqnarray} 
\begin{eqnarray}
  H_{11} &=& \left\{\begin{array}{ll}
\sum_j\Omega_{0;r_1}^{(d_j)} |11\rangle\langle d_j 1|/2 +\text{h.c.},& \text{Pulse-1;}\\
\sum_j\Omega_{1;r_1}^{(d_j)} |r_11\rangle\langle r_1d_j|/2 +\text{h.c.},& \text{Pulse-2;}\\
\sum_j\Omega_{0;r_1}^{(d_j)} |11\rangle\langle d_j 1|/2 +\text{h.c.},& \text{Pulse-3.}
  \end{array}\right. \nonumber
\end{eqnarray} 
In the equations above, we have ignored the interactions of the states $|r_1d_1\rangle$ and $|r_1d_2\rangle$ since they are already much smaller than the detunings of the channels.

The total error of the transformation fidelity is given by
\begin{eqnarray}
 1-\mathcal{F}&=&E_{\text{decay}} +  E_{\text{leak}},\nonumber
\end{eqnarray}
where $E_{\text{leak}}$ can be calculated by using Eqs.~(\ref{Es}) and~(\ref{Eleak}).
There may also be some residual couplings between $|0\rangle$ and some other Rydberg states with large detunings. However, these leaking channels have detunings equal or larger than $\Delta_{d_{1(2)}}\pm \omega_g$, which is again much larger compared with $\omega_g$. In this sense, the leakage for the state $|0\rangle$ mainly comes from $|0\rangle\mapsto|d_0\rangle$ with a detuning of $\omega_g$. This is true for the case of $n=59$, where we have $\Delta_{d_{1}}=-35.2\times2\pi$~GHz and $\Delta_{d_{2}}=33.5\times2\pi$~GHz,  much larger than $\omega_g$.

The numerical results of the gate performance for different $\Omega$'s are plotted in Fig.~\ref{S-02}(b), where the gate time $t_{\text{g}}$, scaled $E_{\text{decay}}$, $E_{\text{leak}}$ and $1-\mathcal{F}$ are presented against $\Omega$. When increasing $\Omega$, we notice that there can be cases where the total fidelity error can be smaller than $10^{-4}$, denoted by the horizontal dashed line. For the case denoted by the solid circle at the right bottom in the inset of Fig.~\ref{S-02}(b), we have a small fidelity error $1-\mathcal{F}=7.2\times10^{-5}$ when $\Omega/2\pi=201$~MHz and $V_{\text{dd}}=\sqrt6\Omega/4=123\times2\pi$~MHz. Importantly, the gate time is only $t_{\text{g}}=10$~ns. Compared with the gate protocols in Ref.~\cite{Theis2016}, where fidelity errors smaller than $10^{-4}$ were also obtained using analytic derivative removal by adiabatic gate pulses, the method here require much smaller $V$, while the method in Ref.~\cite{Theis2016} requires a blockade shift in the range of $(0.7,2.7)\times2\pi$~GHz, as well as the design of the analytic pulse sequence. In principle, the direct dipole-dipole interaction of two nearby Rydberg atoms can be quite large, although one needs to cope with some issues occurred when two Rydberg atoms are too near to each other: their atomic wavefunctions can inevitably overlap~\cite{Xia2013}. On the other hand, the method in Ref.~\cite{Theis2016} is also useful for the protocol here: by using analytic derivative removal by adiabatic gate pulses, our gate then only has the Rydberg state decay as the source of fidelity error. This latter error, as shown by the dotted curve in Fig.~\ref{S-02}(b), can be very small for all the cases.  

If a laboratory can not produce a $\Omega$ as large as the one denoted by the solid circle at the right bottom of the inset in Fig.~\ref{S-02}(b), we can find that for $\Omega/2\pi=7$~MHz, a value that was ever realized for exciting a $59D_{3/2}$ state in Ref.~\cite{Gaetan2009}, our gate is characterized by $t_{\text{g}}=286$~ns and $1-\mathcal{F}=1.47\times10^{-3}$. The reason that we can have a small fidelity error is that here $E_{\text{leak}}=5\times10^{-7}$, thus the decay error is the only source of fidelity error. Here both $\Omega$ and the blockade interaction $V_{\text{dd}}=\sqrt6\Omega/4=4.3\times2\pi$~MHz are within current availability, according to the result in Ref.~\cite{Gaetan2009}. 

An interesting feature in Fig.~\ref{S-02}(b) is that the leakage error $E_{\text{leak}}$ oscillates when $\Omega$ changes~(similar phenomenon exists in Fig.~\ref{fig03}). This can be understood as follows. The rotation error for the state $|0\rangle$ can be from a residue transition $|0\rangle\rightarrow|d_0\rangle$. For a $2\pi$ pulse of the transition $|1\rangle\rightarrow|r_1\rangle$, the rotation error upon the state $|0\rangle$ via $|0\rangle\rightarrow|d_0\rangle$ is thus 
\begin{eqnarray}
  E_{\text{leak,0}} &=&  1-\left|\langle 0|\text{exp}[ -i\hat{H}_{\text{r0}}t_{2\pi}]|0\rangle\right|^2,\nonumber
\end{eqnarray}
where 
\begin{eqnarray}
  \hat{H}_{\text{r0}} &=&[-\omega_g|0 \rangle\langle 0|+ (\Omega' |d_0 \rangle\langle 0|/2 +\text{h.c.} )].
  \label{Hr7}
\end{eqnarray}
Assuming $\Omega'=\Omega$ will give us $E_{\text{leak,0}}\approx2.5\Omega^4/\omega_g^4$ when $\Omega'/\omega_g\ll1$~(as can be easily found by a numerical scaling), which is quite small. However, a small deviation from the condition $\Omega'=\Omega$ will give us another result. We thus want to find an upper bound of this error. This can be done by observing the fact that during the whole Rabi cycle in Eq.~(\ref{Hr7}), the largest population on the level $|d_0\rangle$ is $(\Omega')^2/\omega_g^2$. This means that a small deviation from $\Omega'=\Omega$ may push the error toward $(\Omega')^2/\omega_g^2$. The total effect from the various leaking channels analyzed around Eqs.~(\ref{Hpro1}) results of the complex oscillatory phenomenon of $E_{\text{leak}}$ in Fig.~\ref{S-02}(b). Notice that a similar phenomenon was also reported in Fig.~2 of Ref.~\cite{Theis2016} when a square pulse was used.

%=======================================
%error case 1
%=======================================

With the rapid development of trapping technology of neutral atoms recently~\cite{Reiserer2013,Thompson2013,Kaufman2014,Lester2015}, we may anticipate better trapping of the qubits, so that the intrinsic fidelity error of our gate may reach its intrinsic value in the order of $10^{-5}$ soon.
To conclude, it is possible to realize a rapid and accurate two-qubit $C_Z$ quantum gate with our protocol. It will also be useful to design new functional quantum systems based on our Rydberg gate protocols and other ones~\cite{Muller2009,Carr2013,BhaktavatsalaRao2013,Shi2014,Petrosyan2014,Goerz2014,Paredes-Barato2014,Theis2016Jan,Theis2016,Shi2016-arXiv,Beterov2016}, since whose basic ideas were derived from the well-known Rydberg blockade protocol, a method quite different from that of this work.


%===============================================================================

\end{document}